\documentclass[aps,twocolumn,amsmath,amssymb,superscriptaddress,10pt,floatfix,nofootinbib]{revtex4-2}
\usepackage[dvipsnames]{xcolor}
\usepackage{graphicx}
\usepackage{changepage} % Required for the adjustwidth environment
\usepackage{multirow} % Required for the \multirow command
\usepackage{hyperref}
\usepackage{float}
\usepackage{soul}
\usepackage{lipsum}
\usepackage{ulem}
\usepackage{comment}
\usepackage{orcidlink}
\usepackage[capitalise,noabbrev,nameinlink]{cleveref}
\hypersetup{
citecolor=blue,
 colorlinks=true,
 linkcolor=blue,
 filecolor=magenta,
 urlcolor=cyan,
}

\graphicspath{{./figures/}}

\Crefname{equation}{Eq.}{Eqs.}

\usepackage{titlecaps}
\Addlcwords{a and in and of}
\begin{document}

\title{S-wave kaon condensation in neutron-star matter within a chiral model framework with dynamical meson masses
}

\author{Yuhan Wang}
\affiliation{Center for Nuclear Research, Department of Physics, Kent State University, Kent, OH  44242, USA}

\author{Rajesh Kumar\,\orcidlink{0000-0003-2746-3956}}
\affiliation{Department of Physics, Kent State University, Kent, OH 44243, USA}
\affiliation{Department of Physics, MRPD Government College Talwara, Punjab, 144216, India}

\author{Joaquin Grefa}
\affiliation{Center for Nuclear Research, Department of Physics, Kent State University, Kent, OH  44242, USA}
\affiliation{Department of Physics, University of Houston, Houston, TX 77204, USA}

\author{Konstantin Maslov}
\affiliation{Department of Physics, University of Houston, Houston, TX 77204, USA}

\author{Claudia Ratti}
\affiliation{Department of Physics, University of Houston, Houston, TX 77204, USA}

\author{Rodrigo Negreiros \orcidlink{0000-0002-9669-905X}}
%\email{rnegreiros@catholic.tech}
\affiliation{Department of Physics, Catholic Institute of Technology, MA, USA}
\affiliation{Department of Physics, Universidade Federal Fluminense, Niteroi, Brazil}
\affiliation{
ICRANet, Piazza della Repubblica 10, I-65122 Pescara, Italy}

\author{Veronica Dexheimer}
% \email{vdexheim@kent.edu}
\affiliation{Center for Nuclear Research, Department of Physics, Kent State University, Kent, OH  44242, USA}

\date{\today}

\begin{abstract}
We investigate s-wave kaon condensation in dense matter and neutron stars within the updated Chiral Mean Field model with an improved meson description (mCMF), which incorporates dynamically generated in-medium meson masses arising from explicit chiral symmetry breaking and vector-meson self-interactions. In contrast to conventional relativistic mean-field descriptions with constant meson masses, the mCMF framework introduces a self-consistent feedback between the meson sector and the dense-matter equations of motion. 
The kaon dispersion relation is derived from the nonlinear SU(3) Lagrangian, including the Weinberg--Tomozawa interaction and additional baryon--pseudoscalar couplings, and the onset of condensation is determined under conditions of charge neutrality and $\beta$ equilibrium. Our calculations include the full baryon octet together with electrons and muons at zero temperature. 
We analyze the impact of hyperons, muons, and kaon condensation on the equation of state, on neutron-star mass--radius relations, and neutron-star thermal evolution, and examine the sensitivity of the onset density and stellar properties to variations in the nucleon--kaon scattering length and to different model vector parameters and vector self-interactions. We find that $K^{-}$ condensation sets in between $n \sim (2-8)\, n_0$ (in units of nuclear saturation density) and leads to a moderate to strong softening (in one case, a slight stiffening of the equation of state), depending on the interplay of kaons and hyperons, while remaining compatible with current $2\,M_\odot$ and small-radius neutron-star observational constraints and producing distinguishable behavior in the neutron-star cooling. This work provides an improved and thermodynamically consistent framework for studying exotic degrees of freedom in neutron-star matter. 
\end{abstract}

\maketitle
%\tableofcontents
%\listoffigures

\section{Introduction}

Understanding the properties of strongly interacting matter at densities several times nuclear saturation density remains one of the central challenges in nuclear astrophysics. Neutron stars (NSs), born in core-collapse supernovae, provide a unique laboratory to probe high-energy matter under extreme conditions of density and isospin asymmetry that cannot be reproduced in terrestrial experiments. Observations of massive pulsars, including PSR J1614$-$2230~\cite{Demorest:2010bx}, PSR J0348$+$0432~\cite{Antoniadis:2013pzd}, and PSR J0740$+$6620~\cite{Fonseca:2021wxt}, with masses around $2\,M_\odot$, impose stringent constraints on the equation of state (EoS) of dense matter. Any viable theoretical model must reconcile the possible appearance of exotic degrees of freedom with these astrophysical constraints, in addition to data related to the dynamical evolution of neutron stars.

At sufficiently high baryon chemical potentials, additional hadronic or non-hadronic degrees of freedom are expected to emerge. The inclusion of hyperons generally softens the EoS due to the opening of new Fermi channels, potentially reducing the maximum neutron star mass below the observed $2\,M_\odot$ limit, especially when radius or tidal deformability constraints are also imposed. This problem is commonly referred to as the hyperon puzzle~\cite{Chamel:2013efa}, where hyperons reduce the possible maximum mass of neutron stars. Muons are associated with a similar effect (although the main difference now is in the stellar radius, not so much the mass, and it depends on the symmetry energy and its slopes~\cite{1988PhRvC..38.1010W,Zhang:2020wov}). Meson condensates may also appear in dense matter. Although they are bosonic rather than fermionic, their appearance can induce similar or even stronger softening effects due to phase transitions in dense matter. In particular, the possibility of s-wave antikaon ($K^{-}$) condensation has attracted considerable attention since the pioneering work of Kaplan and Nelson~\cite{Kaplan:1986yq,Kaplan:1987sc}. There, they demonstrated (within a nonlinear realization of chiral symmetry) that attractive kaon--nucleon interactions could lower the in-medium kaon energy sufficiently to trigger Bose--Einstein condensation when the electron chemical potential exceeds the kaon energy~\cite{Kaplan:1986yq,Kaplan:1987sc}.

Subsequent studies examined the implications of $K^{-}$ condensation for neutron-star structure, including the possibility of a first-order transition and the formation of mixed phases~\cite{Thorsson:1993bu,Glendenning:1997ak}. The competition and coexistence of hyperons and antikaon condensation were later explored within relativistic mean-field and chiral-based approaches, where the appearance of additional hadronic degrees of freedom can substantially shift the condensation threshold and modify the stellar composition and EoS stiffness (see recent work in Refs.~\cite{Banik:2001yw,Lee:1994jj,Thapa:2021kfo,Malik:2021nas,Muto:2021jms,Ma:2022fmu,Muto:2022ces,Kundu:2022nva,Ma:2022knr,Kaur:2024cfm,Muto:2024upf,Muto:2025jaq,S:2025xfm,GuhaRoy:2025kht}). These works established that the onset density of kaon condensation is highly sensitive to the antikaon optical potential (or equivalently the in-medium kaon self-energy), to low-energy kaon--nucleon interaction parameters, and to the presence of hyperons. However, many existing implementations either adopt relativistic mean-field descriptions with constant meson masses and couplings, or restrict attention to threshold conditions and simplified treatments of the condensed phase, without a self-consistent description of the condensed phase within a unified model capable of producing realistic neutron-star EoSs.  

Chiral symmetry provides a systematic framework to describe in-medium kaon dynamics. In chiral effective models, the kaon dispersion relation in baryonic matter receives characteristic contributions from the Weinberg--Tomozawa interaction, scalar meson mean fields, and explicit chiral symmetry breaking, which together determine the in-medium kaon self-energy and the associated optical potential~\cite{Brown:1993yv,Schaffner:1995th}. In-medium modifications of kaon properties---including effective masses and dispersion relations---have been investigated in nonlinear SU(3) sigma-model extensions and related approaches~\cite{Mishra:2004te,Mishra:2006wy,Kumar:2020vys}. Nevertheless, a fully self-consistent treatment of $K^{-}$ condensation embedded in an EoS framework that simultaneously (i) reproduces realistic neutron-star matter under $\beta$ equilibrium with the full baryon octet and leptons and (ii) incorporates dynamically generated in-medium meson masses that feed back into the mean-field equations remains unexplored.

In this work, we investigate s-wave kaon condensation at zero temperature in dense matter within the updated hadronic Chiral Mean Field model with an improved meson description (mCMF). The mCMF framework extends the original Chiral Mean Field (CMF) model by incorporating dynamical in-medium meson masses arising from explicit chiral symmetry breaking in the pseudoscalar sector and vector meson self-interactions. In contrast to conventional RMF models where meson masses are fixed parameters, in mCMF the pseudoscalar and vector meson masses depend explicitly on the scalar and vector meson mean fields, respectively, introducing a nontrivial feedback mechanism into the dense-matter equations of motion.

The kaon dispersion relation is obtained by Fourier transforming the equations of motion derived from the nonlinear SU(3) Lagrangian, including the Weinberg--Tomozawa interaction, the first range term that includes a coupling between scalar and pseudoscalar mesons in addition to the pseudoscalar kinetic energy term, and additional baryon--pseudoscalar derivative couplings characterized by the $d_1$ and $d_2$ parameters. The onset of s-wave $K^{-}$ condensation is determined by the zero-momentum condition $\omega_{K^-}(k=0) = \mu_e$, where $\omega_{K^-}$ is the in-medium kaon energy and $\mu_e$ is the electron chemical potential. We then, for the first time, solve self-consistently the extended set of mean-field equations including the kaon condensate under conditions of charge neutrality and $\beta$ equilibrium. The other kaons do not condense under the conditions we investigate.

Our calculations include the full baryon octet together with electrons and muons. We systematically analyze the impact of hyperons, muons, and kaon condensation on the EoS, on neutron-star mass--radius relations, and on neutron-star thermal evolution.
Since the first investigation of the effect of kaon condensation on the thermal evolution of neutron stars~\cite{1990ApJ...354L..17P}, the main question addressed has been how kaon condensation affects the different neutrino emission processes in neutron stars and whether this can be used as a way to distinguish stars with kaon condensation in their interiors.

In this work, we investigate the sensitivity of the kaon condensation onset density and stellar properties to variations in the nucleon--kaon scattering length through the $d_1$ and $d_2$ couplings. We also study different vector parameters and vector self-interactions. This allows us to quantify uncertainties in how interactions propagate to macroscopic neutron-star observables.
By embedding kaon condensation within a framework that already reproduces nuclear matter properties, satisfies astrophysical $\sim2\,M_\odot$ constraints, and incorporates dynamically generated in-medium meson masses, the present study provides a unified and thermodynamically consistent treatment of exotic degrees of freedom in neutron-star matter. 

\section{Formalism}

\subsection{Standard mCMF Lagrangian}

We provide a comprehensive review of the CMF model in~Ref.~\cite{Cruz-Camacho:2024odu}. Here we summarize our main equations starting from the Lagrangian density
\begin{align}
\mathcal{L}_{\rm CMF} = \mathcal{L}_{\rm kin} + \mathcal{L}_{\rm int} + \mathcal{L}_{\rm scal} + \mathcal{L}_{\rm vec} + \mathcal{L}_{m_0} + \mathcal{L}_{\rm esb}\,,
\label{all}
\end{align}
where \( \mathcal{L}_{\rm kin} \) represents the kinetic terms for the baryon octet and leptons, \( \mathcal{L}_{\rm int} \) the baryon interaction with the scalar ($\sigma$, $\zeta$, and $\delta$) and vector ($\omega$, $\phi$, and $\rho$) mean-field mesons, \(\mathcal{L}_{\rm scal}\) the self-interactions of scalar mesons, \( \mathcal{L}_{\rm vec} \) the vector meson masses and self-interactions, and \( \mathcal{L}_{\rm esb} \) the explicit chiral symmetry breaking. The $\mathcal{L}_{m_0}$ term is added in order to fit the compressibility for the C4 and RC4 vector couplings used in this work. Finally, we subtract the constant mean-field term $\mathcal{L}^{\rm M. F.}_{\rm {vacuum}}$ to reproduce zero vacuum pressure. 

The interaction with the vector mean-field mesons gives rise to an effective effective chemical potential for baryon $i$
\begin{gather}
    \mu_i^*=\mu_i-g_{\omega i}\omega_0-g_{\rho i}\rho_0-g_{\phi i}\phi_0\,,
\end{gather}
with couplings $g_{i}$ to the respective vector mesons and particle $i$ (baryon, meson, or lepton) chemical potential written as
\begin{align}
\mu_i=B_i\mu_B+Q_i\mu_Q\,,
\label{mu}
\end{align}
where $B_i$ and $Q_i$ are the particle $i$'s baryon number and electric charge (the two conserved quantities in the system), and $\mu_B$ and $\mu_Q$ are the respective chemical potentials. For leptons and mesons, $\mu_i^*=\mu_i$ since no interactions of mesons with vector meson mean fields are present in our approach, and leptons do not interact via the strong force. 

The effective or in-medium mass for the baryons is
\begin{align}
m_i^*=g_{\sigma i}\sigma+g_{\zeta i}\zeta+g_{\delta i}\delta+\Delta m_i\,,
\label{eq:emh}
\end{align}
with couplings $g_{i}$ with the respective scalar mesons. For leptons $m_i^*=m_i$. The parameter $\Delta m_i$ accounts for the bare mass $m_0$ associated with the baryon octet and additionally the explicit symmetry-breaking parameter $m^{HO}_3$ for the hyperons only.

The vector term from the Lagrangian can explicitly be written for the C4 parametrization~\cite{Cruz-Camacho:2024odu} as 
\begin{align}
&\mathcal{L}_{\rm vec}^{(C4)}=\frac{1}{2}\left(m_\omega^2\omega^2+m_\phi^2\phi^2+m_\rho^2\rho^2\right)\nonumber \\
&+g_4\left(\omega^4+2\sqrt2\omega^3\phi+3\omega^2\phi^2+\sqrt2\omega\phi^3+\frac{\phi^4}{4}\right)\,, 
\label{eq:L_vec}
\end{align}
while the RC4 parametrization involves a redefinition of
the vector meson fields, which enables precise adjustments
of meson masses~\cite{Kumar:2024owe}
\begin{align}
\mathcal{ L}_{\rm vec}^{RC4}&=
\frac{1}{2} \left(m_\rho^2 \rho^2+m_\omega^2 \omega^2+m_\phi^2 \phi^2\right)\nonumber\\&+
g_4\Bigg(\omega^{4}+2\sqrt{2}\bigg(\frac{Z_\phi}{Z_\omega} \bigg)^{1/2} \omega^{3} \phi+3 \bigg(\frac{Z_\phi }{Z_\omega}\bigg)  \omega^{2} \phi^{2}\nonumber\\&+
\sqrt{2} \bigg(\frac{Z_\phi}{Z_\omega}\bigg)^{3/2}\omega \phi^{3}+\frac{1}{4} \bigg(\frac{Z_\phi }{Z_\omega}\bigg)^2\phi^{4}\Bigg),
\label{eq:L_vec_kin_m}
\end{align}
where $Z_\phi$ and $Z_\omega$ are fitted to reproduce different vacuum masses for the $\rho$ (same as $\omega$) and $\phi$ vector mesons. In principle, all the $g$ couplings between vector mesons and baryons (together with the $g_4$ coupling) also need to be rescaled. Nevertheless, since they are fitted to  data, we can simply perform a new fit. See tables in Ref.~\cite{Kumar:2024owe} for RC4 and in Ref.~\cite{Cruz-Camacho:2024odu} for C4  for the values of the corresponding parameters and couplings. CMF parameters are adjusted to reproduce particle vacuum properties, saturation nuclear properties, neutron star observables, and (for the quark sector) lattice QCD results.
Without redefinition, either all vacuum masses of the vector mesons are taken to be the same or, as in the standard C4 parametrization, they are taken to be different, but this slightly breaks the symmetry that gives rise to all the terms in the CMF model. 

For this work, we include an additional vector-isovector interaction $g_{\omega\rho}\omega^2\rho^2$~\cite{Horowitz:2002mb} to \Cref{eq:L_vec_kin_m}. It has been shown to lead to neutron star masses and tidal deformabilities in better agreement with observations~\cite{Dexheimer:2018dhb}. In the $C4$ parametrization, it has coupling $g_{\omega\rho}/g_{N\omega}=135/11.9=11.34$ (normalized to the nucleon-$\omega$ coupling), while for the $RC4$ parametrization we use it for the first time with normalized coupling $g_{\omega\rho}/g_{N\omega}=135/11.8= 11.44$. The $g_{\omega\rho}/g_{N\omega}$ value was chosen to reproduce the tidal deformability $\Lambda< 730$ for C4, in agreement with results obtained from the binary neutron-star merger GW170817~\cite{LIGOScientific:2018hze}. For both couplings, $g_{N\rho}$ is then fitted to reproduce the same symmetry energy at saturation as for the respective case without the $\omega\rho$ interaction.

In addition to the mean-field standard CMF explicit symmetry breaking term in \Cref{all}, mCMF has additional terms for the pseudoscalar mesons. Ignoring the vector and other pseudoscalar thermal mesons described in mCMF~\cite{Kumar:2025rxj} and keeping only kaons, the explicit symmetry breaking term results in
\begin{align} 
\mathcal{L}_{\rm esb}^{\rm{mCMF}}&=\frac{m_K^2}{2 f_K}\Big[(\sigma+\sqrt{2} \zeta+\delta)\left(K^{+} K^{-}\right)\nonumber\\&+(\sigma+\sqrt{2} \zeta-\delta)\left(K^0 \bar{K}^0\right)\Big]\,,
\label{7}
\end{align}
where $m^2_{K}=498$ MeV is the vacuum mass of the kaons and $\sigma_{0}$ and $\zeta_{0}$ are the vacuum values of the scalar mesons; $\delta_{0}=0$.

In the non-linear realization of chiral symmetry, the pseudoscalar mesons are the parameters of the chiral symmetry transformation. To derive the Lagrangian for mCMF, we expand the chiral symmetry transformation operator $u$ to second order in the pseudoscalar meson fields and then calculate the kaon masses before applying the mean-field approximation
\begin{align}
m^{*}{_{K}^2} &=\lim_{K \rightarrow \langle K \rangle}\frac{\partial^2 U}{\partial K^2} \,,
\label{eq:meson_mass_formula}
\end{align}
where $U$ is (minus) the potential part of the Lagrangian (omitting the constant vacuum term) 
\begin{align}
U = - \mathcal{L}_{\rm int} - \mathcal{L}_{\rm scal} - \mathcal{L}_{\rm vec} - \mathcal{L}_{m_0} - \mathcal{L}_{\rm esb}-\mathcal{L}_{\rm esb}^{\rm{mCMF}}\,,
\end{align}
where only the last term contributes to the derivative and the vacuum expectation of mesons is taken to be $\left<K\right>=0$. The kaon effective masses then read
\begin{align} 
{m^{*^2}_{K^+/K^-}}&= \frac{0.5 m^2_{K}  \left(2 \zeta + \sqrt{2} \left(\delta + \sigma\right)\right) \left(\sqrt{2} \sigma_{0} + 2 \zeta_{0}\right)}{\left(\sigma_{0} + \sqrt{2} \zeta_{0}\right)^{2}}\,,
\label{eq:non_deg_Kpm_mass}
\\
{m^{*^2}_{K^0/\bar K^0}}&=\frac{0.5  m^2_{K}  \left(2 \zeta + \sqrt{2} \left(-\delta + \sigma\right)\right) \left(\sqrt{2} \sigma_{0} + 2 \zeta_{0}\right)}{\left(\sigma_{0} + \sqrt{2} \zeta_{0}\right)^{2}}\,.
\label{eq:non_deg_K0b_mass}
\end{align}

Since the kaon masses depend on the scalar mesons, they feed back into the scalar meson equations of motion. We discuss the exact terms this such mechanism generates in the following. We note that at zero temperature, the thermal meson contributions discussed in Ref.~\cite{Kumar:2025rxj} do not contribute. Notice that mCMF describes in-medium masses for vector and pseudoscalar thermal mesons. In this work, we include only mesons from the pseudoscalar nonet (not the vector meson $K^*$, which is a kaon resonance).

\subsection{Complete Lagrangian for kaons}

The interaction Lagrangian density describing the kaon and antikaon interaction with the baryons and with the scalar mesons that we use in this work (going before the previously discussed mCMF description) is given by~\cite{Mishra:2004te} 
\begin{widetext}
\begin{align} \mathcal{L}_{\rm{kaon}}= & -\frac{i}{4 f_K^2}\left[\left(2 \bar{p} \gamma^\mu p+\bar{n} \gamma^\mu n-\bar{\Sigma} ^{-}\gamma^\mu \Sigma^{-}+\bar{\Sigma}^{+} \gamma^\mu \Sigma^{+}-2 \bar{\Xi}^{-} \gamma^\mu \Xi^{-}-\bar{\Xi}^0 \gamma^\mu \Xi^0\right)\left(K^{-}\left(\partial_\mu K^{+}\right)-\left(\partial_\mu K^{-}\right) K^{+}\right)\right. \nonumber \\ & \left.+\left(\bar{p} \gamma^\mu p+2 \bar{n} \gamma^\mu n+\bar{\Sigma} ^{-}\gamma^\mu \Sigma^{-}-\bar{\Sigma} ^{+}\gamma^\mu \Sigma^{+}-\bar{\Xi}^{-} \gamma^\mu \Xi^{-}-2 \bar{\Xi}^0 \gamma^\mu \Xi^0\right)\left(\bar{K}^0\left(\partial_\mu K^0\right)-\left(\partial_\mu \bar{K}^0\right) K^0\right)\right]\nonumber\\ &+\frac{m_K^2}{2 f_K}\left[(\sigma+\sqrt{2} \zeta+\delta)\left(K^{+} K^{-}\right)+(\sigma+\sqrt{2} \zeta-\delta)\left(K^0 \bar{K}^0\right)\right]\nonumber \\ & -\frac{1}{f_K}\left[(\sigma+\sqrt{2} \zeta+\delta)\left(\partial_\mu K^{+}\right)\left(\partial^\mu K^{-}\right)+(\sigma+\sqrt{2} \zeta-\delta)\left(\partial_\mu K^0\right)\left(\partial^\mu \bar{K}^0\right)\right]\nonumber \\ & +\frac{d_1}{2 f_K^2}\left(\bar{p} p+\bar{n} n+\bar{\Lambda}^0 \Lambda^0+\bar{\Sigma}^{+} \Sigma^{+}+\bar{\Sigma}^0 \Sigma^0+\bar{\Sigma}^{-} \Sigma^{-}+\bar{\Xi}^{-} \Xi^{-}+\bar{\Xi}^0 \Xi^0\right)\left(\left(\partial_\mu K^{+}\right)\left(\partial^\mu K^{-}\right)+\left(\partial_\mu K^0\right)\left(\partial^\mu \bar{K}^0\right)\right)\nonumber \\ & +\frac{d_2}{2 f_K^2}\left[\left(\bar{p} p+\frac{5}{6} \bar{\Lambda}^0 \Lambda^0+\frac{1}{2} \bar{\Sigma}^0 \Sigma^0+\bar{\Sigma}^{+}{\Sigma^{+}}^{-} \bar{\Xi}^{-} \Xi^{-}+\bar{\Xi}^0 \Xi^0\right)\left(\partial_\mu K^{+}\right)\left(\partial^\mu K^{-}\right)\right.\nonumber \\ & \left.+\left(\bar{n} n+\frac{5}{6} \bar{\Lambda}^0 \Lambda^0+\frac{1}{2} \bar{\Sigma}^0 \Sigma^0+\bar{\Sigma}^{-} \Sigma^{-}+\bar{\Xi}^{-} \Xi^{-}+\bar{\Xi}^0 \Xi^0\right)\left(\partial_\mu K^0\right)\left(\partial^\mu \bar{K}^0\right)\right])\,. 
\label{eq:Ldensity}
\end{align}
\end{widetext} 
The first term in~\Cref{eq:Ldensity} is the so-called the Weinberg–Tomozawa term (minus the baryon kinetic term). It corresponds to a vector interaction, but while it is repulsive for $K$ mesons ($K^+$ and $K^0$), it is attractive for $\bar{K}$ mesons (the respective antiparticles, $K^-$ and $\bar{K}^0$). The coefficient $f_K = 122~ \mathrm{MeV}$ denotes the kaon decay constant. 
The second term originates from the explicit chiral symmetry-breaking term and leads to the scalar meson exchange interaction (already in mCMF~\cite{Kumar:2025rxj} and described in \Cref{7}). The third term is called the first range term. The last two terms with parameters $d_1$ and $d_2$ are the contact terms. 
This Lagrangian term contains derivative couplings, which were not previously included in mCMF, where the pseudoscalar meson in-medium masses were calculated before the mean-field approximation.

The coefficients $d_1$ and $d_2$ are fixed from the kaon--nucleon scattering lengths by inverting the relations
\begin{align}
a_{K N}(I=0)= & \frac{m_K}{4 \pi f_K^2\left(1+m_K / m_N\right)} \nonumber \\&
\times\left[\frac{m_K f_K}{2}\left(\frac{g_{\sigma N}}{m_\sigma^{2}} +\sqrt{2}\frac{g_{\zeta N}}{m_\zeta^{2}}-3\frac{g_{\delta N}}{m_\delta^{2}}\right)\right.\nonumber \\&
\left.+\frac{\left(d_1-d_2\right) m_K}{2}\right]\,,
\label{e1}
\end{align}
and
\begin{align}
a_{K N}(I=1)= & \frac{m_K}{4 \pi f_K^2\left(1+m_K / m_N\right)}\nonumber \\
& \times\left[-1+\frac{m_K f_K}{2}\left(\frac{g_{\sigma N}}{m_\sigma^{2}}+\sqrt{2}\frac{g_{\zeta N}}{m_\zeta^{2}}+\frac{g_{\delta N}}{m_\delta^{2}}\right)\right.\nonumber \\
& \left.+\frac{\left(d_1+d_2\right) m_K}{2}\right]\,,
\label{e2}
\end{align}
for different isospin channels $I=0$ and $I=1$. These follow from the  matrix amplitude ($T$) evaluated at threshold through
\begin{equation}
a_{KN}=\frac{1}{4\pi\left(1+\frac{m_K}{m_N}\right)}\,T,
\end{equation}
with  where $T$ is the tree-level $KN\to KN$ invariant amplitude at threshold, computed from $\mathcal{L}_{\rm kaon}$ (Eq.~12) and the baryon--scalar couplings in $\mathcal{L}_{\rm int}$~\cite{Barnes:1992ca}.

Experimental and theoretical determinations of kaon--nucleon scattering lengths exhibit a considerable spread. 
For example, scattering data for $K^{+}p$ typically yield values of the real part in the range $a\approx -1.05$ to $-0.31~\rm fm$ \cite{ALICE:2021szj,Doring:2011xc}. 
For the $K^{-}p$ channel, analyses of scattering and kaonic atom data give values around $a\approx -0.65~\rm fm$ \cite{Mai:2012dt}, while theoretical calculations within chiral coupled-channel approaches obtain values such as $a(K^-p)\approx -1.05~\rm fm$ and $a(K^-n)\approx 0.57~\rm fm$ \cite{Borasoy:2006sr,Ikeda:2012au}. 
These can be translated into the isospin basis through
\begin{equation}
a_p=\frac12 a(I=0)+\frac12 a(I=1), \qquad a_n=a(I=1),
\end{equation}
leading to typical ranges $a(I=0)\approx -1.81$ to $-1.23~\rm fm$ and $a(I=1)\approx -0.06$ to $0.48~\rm fm$ \cite{Mai:2012dt,Meissner:2006gx,Meissner:2004jr,Borasoy:2006sr}. 
At low energy, a positive scattering length indicates an attractive interaction capable of supporting a bound state, while a negative value corresponds either to a weaker attractive interaction, although not strong
enough to form a bound state, or to a repulsive interaction.

Given the spread among existing determinations, we adopt as a reference the values commonly used in studies of kaon condensation in dense matter, $a_{KN}(I=0)=-0.09~\rm fm$ and $a_{KN}(I=1)=-0.31~\rm fm$ \cite{Barnes:1992ca}. 
These values lead to the parameters
\begin{equation}
d_1=\frac{2.30}{m_K}, \qquad d_2=\frac{0.71}{m_K},
\end{equation}
in terms of the kaon mass. 
However, the kaon--nucleon interaction relevant for $K^{-}$ condensation in dense matter is not completely constrained by threshold scattering data alone. 
In chiral approaches, the $K^-p$ amplitude is strongly influenced by the nearby $\Lambda(1405)$ resonance, while the $K^-n$ interaction exhibits a smoother energy dependence \cite{Lee:1994my}. 
Moreover, the effective $\bar K N$ interaction is coupled to channels such as $\pi\Sigma$, implying that the interaction strength entering the in-medium kaon self-energy may differ from that inferred directly from threshold scattering lengths \cite{Hyodo:2007jq}. 
Since the onset of kaon condensation depends sensitively on the kaon self-energy and therefore on the underlying kaon--nucleon interaction strength, uncertainties in these quantities can significantly affect the predicted condensation density.

For this reason, in addition to the reference parameter set, we perform an exploratory variation of the contact terms $d_1$ and $d_2$. 
In particular, we scale the reference values up to $9d_1$ and $9d_2$ in order to assess the sensitivity of the condensation threshold and neutron-star observables to a stronger effective kaon--baryon interaction. 
Within Eqs.~(\ref{e1}) and (\ref{e2}), this choice corresponds to effective scattering lengths $a(I=0)=2.09~\rm fm$ and $a(I=1)=3.82~\rm fm$. 
We emphasize that this enlarged range should be interpreted as an exploratory sensitivity analysis rather than as a direct determination of the density dependence of the physical kaon--nucleon scattering lengths. Finally, note that due to the complicated form of \Cref{e1,e2}, even using $a_{K N}(I=0)=0$ and $a_{K N}(I=1)=0$ corresponds to non-zero parameters $d_1=3.47\,\rm{fm}$ and $d_2=1.35\,\rm{fm}$ (which still allows for meson condensation).

\subsection{Dispersion relation for kaons}

We write the dispersion relation for each kaon species in terms of energy $\omega$, momentum $\vec{k}$ and mass, including the (in-medium) self-energy $\Pi^*$ 
\begin{align}  
-\omega^2+\vec{k}^2+m_{K}^2-\Pi^*(\omega,|\vec{k}|)=0\,.
\label{eq:disp}
\end{align}

The self-energy for each kaon is derived from the Fourier transform of the respective equation of motion. For kaons $K$ ($K^+\rm{or\,K^0}$) and antikaons $\bar{K}$ (${K^-\rm{or\,}\bar K^0}$) they are
\begin{widetext}
\begin{align}
\Pi_{K}^*= & -\frac{1}{4 f_K^2}\left[3\left(n_p^v+n_n^v\right) \pm\left(n_p^v-n_n^v\right) \pm 2\left(n_{\Sigma^{+}}^v-n_{\Sigma^{-}}^v\right)-\left[3\left(n_{\Xi^{-}}^v+n_{\Xi^0}^v\right) \pm\left(n_{\Xi^{-}}^v-n_{\Xi^0}^v\right)\right]\right] \omega
+\frac{m_K^2}{2 f_K}\left(\sigma^{\prime}+\sqrt{2} \zeta^{\prime} \pm \delta^{\prime}\right)\nonumber \\
& +\Bigg[-\frac{1}{f_K}\left(\sigma^{\prime}+\sqrt{2} \zeta^{\prime} \pm \delta^{\prime}\right)+\frac{d_1}{2 f_K^2}\left(n_p^s+n_n^s+n_{\Lambda^0}^s+n_{\Sigma^{+}}^s+n^s_{\Sigma^0}+n_{\Sigma^{-}}^s+n_{\Xi^{-}}^s+n^s_{\Xi^0}\right)\nonumber \\ & +\frac{d_2}{4 f_K^2}\left(\left(n_p^s+n^s_n\right) \pm\left(n_p^s-n^s_n\right)+n^s_{\Sigma^0}+\frac{5}{3} n^s_{\Lambda^0}+\left(n^s_{\Sigma^{+}}+n^s_{\Sigma^{-}}\right) \pm\left(n^s_{\Sigma^{+}}-n^s_{\Sigma^{-}}\right)\right.
+2 n^s_{\Xi^{-}}+2 n^s_{\Xi^0}\Bigg)\Bigg]\left(\omega^2-\vec{k}^2\right),
\\
\Pi_{\bar{K}}^*= & \frac{1}{4 f_K^2}\left[3\left(n_p^v+n_n^v\right) \pm\left(n_p^v-c\right) \pm 2\left(n_{\Sigma^{+}}^v-n_{\Sigma^{-}}^v\right)-\left[3\left(n_{\Xi^{-}}^v+n_{\Xi^0}^v\right) \pm\left(n_{\Xi^{-}}^v-n_{\Xi^0}^v\right)\right]\right] \omega+\frac{m_{\bar{K}}^2}{2 f_{\bar{K}}}\left(\sigma^{\prime}+\sqrt{2} \zeta^{\prime} \pm \delta^{\prime}\right)\nonumber \\
& +\Bigg[-\frac{1}{f_{\bar{K}}}\left(\sigma^{\prime}+\sqrt{2} \zeta^{\prime} \pm \delta^{\prime}\right)+\frac{d_1}{2 f_{\bar{K}}^2}\left(n_p^s+n_n^s+n_{\Lambda^0}^s+n_{\Sigma^{+}}^s+n_{\Sigma^0}^s+n_{\Sigma^{-}}^s+n_{\Xi^{-}}^s+n_{\Xi^0}^s\right) \nonumber\\
& +\frac{d_2}{4 f_K^2}\Bigg(\left(n_p^s+n_n^s\right) \pm\left(n_p^s-n_n^s\right)+n_{\Sigma^0}^s+\frac{5}{3} n_{\Lambda^0}^s+\left(n_{\Sigma^{+}}^s+n_{\Sigma^{-}}^s\right) \pm\left(n_{\Sigma^{+}}^s-n_{\Sigma^{-}}^s\bigg) +2 n_{\Xi^{-}}^s+2 n_{\Xi^0}^s\right)\Bigg]\left(\omega^2-\vec{k}^2\right),
\end{align}
\end{widetext}
where the following rescaling by vacuum values was performed for the scalar mesons: $\sigma^{\prime}=(\sigma-\sigma_0)$, $\zeta^{\prime}=(\zeta-\zeta_0)$, $\delta^{\prime}=(\delta-\delta_0)$ and  $n^v$ and $n^s$ represent the standard vector and scalar (number) densities for the baryons. 

The in-medium effective masses for the kaons are now calculated from the dispersion relation \Cref{eq:disp} (instead of how they were calculated for mCMF using \Cref{eq:meson_mass_formula}). Explicitly, for zero momentum, we solve numerically 
\begin{align}  
\omega^2=m_{K}^2-\Pi^*(\omega)\,.
\label{eq:disp}
\end{align} 

\subsection{Vector density for kaons}
\label{2D}

The conserved current associated with kaons is obtained from  the relevant Lagrangian \Cref{eq:Ldensity} as 
\begin{align}
J^{\mu}_{K}=i\left(\bar{K}\frac{\partial\mathcal{L}}{\partial(\partial_{\mu}\bar{K})}-\frac{\partial\mathcal{L}}{\partial(\partial_{\mu}K)}K\right) \,,
\end{align}

The vector density or time component of the current (hereon simply referred to as kaon density) for the $K^-$ is (from \cite{Kumari:2022jvq} correcting for a minus sign)
\begin{widetext}
\begin{align}
&n_{{K^-}/{K^+}}=-J_{K^{-}/{K^+}}^0
=\pm \left\{ \frac{1}{2 f_K^2} \left( 2 n_p^v + n_n^v - n_{\Sigma^{-}}^v + n_{\Sigma^{+}}^v - 2 n_{\Xi^{-}}^v - n_{\Xi^0}^v \right) \right.
\nonumber + 2 \omega \left[ 1 - \frac{1}{f_K} \left( \sqrt{2} \zeta^{\prime} + \delta^{\prime}+\sigma^{\prime} \right) \right. \nonumber \\
& + \frac{d_1}{2 f_K^2} \left( n_p^s + n_n^s + n_{\Lambda^0}^s + n_{\Sigma^{+}}^s + n_{\Sigma^0}^s + n_{\Sigma^{-}}^s + n_{\Xi^{-}}^s + n_{\Xi^0}^s \right) 
\left. \left. + \frac{d_2}{2 f_K^2} \left( n_p^s + \frac{5}{6} n_{\Lambda^0}^s + \frac{1}{2} n_{\Sigma^0}^s + n_{\Sigma^{+}}^s + n_{\Xi^{-}}^s + n_{\Xi^0}^s \right) \right] \right\} K^{-} K^{+}\,.
\\
&n_{\bar{K^{0}}/{K^0}}=-J^0_{\bar{{K}^{0}}//{K^0}}=\pm \left\{ \frac{1}{2 f_K^2} \left(  n_p^v +2 n_n^v - n_{\Sigma^{-}}^v + n_{\Sigma^{+}}^v - n_{\Xi^{-}}^v -2 n_{\Xi^0}^v \right) \right.+ 2 \omega \left[ 1 - \frac{1}{f_K} \left( \delta^{\prime} + \sqrt{2} \zeta^{\prime} - \sigma^{\prime} \right) \right. \nonumber \\
&+ \frac{d_1}{2 f_K^2} \left( n_p^s + n_n^s + n_{\Lambda^0}^s + n_{\Sigma^{+}}^s + n_{\Sigma^0}^s + n_{\Sigma^{-}}^s + n_{\Xi^{-}}^s + n_{\Xi^0}^s \right)\left. \left. + \frac{d_2}{2 f_K^2} \left( n_n^s + \frac{5}{6} n_{\Lambda^0}^s + \frac{1}{2} n_{\Sigma^0}^s + n_{\Sigma^{-}}^s + n_{\Xi^{-}}^s + n_{\Xi^0}^s \right) \right] \right\} \bar{K^{0}}K^{0},
\end{align}    
\end{widetext}
where $|K|=(K^{-} K^{+})^{1 / 2}$ denotes the amplitude of the kaon field, and the angle (also referred to as the rotation angle, chiral angle, or the mixing angle) is $\theta=\sqrt{2}|K|/f_K$. We come back to this angle in the context of thermal evolution of neutron stars in the following.
While above we used the notation $n^s$ for scalar density and $n^v$ for vector density, for simplicity, hereafter we simply use $n$ for the vector density.
As discussed in the following in \Cref{results}, the number density for the other kaons remains zero since those mesons do not condense and, as a result, do not appear in our formalism for the conditions evaluated in this work.

\section{Kaon Condensation}

\subsection{Charge neutrality}

Before the onset of kaon condensation, charge neutrality is maintained by a balance between charged baryons and leptons
\begin{align}
n_p
+n_{\Sigma^+}
-n_{\Sigma^-}
-n_{\Xi^-}
-n_e
-n_\mu=0\,,
\end{align}
After the kaon onset, $K^-$ condenses and starts to contribute to the charge balance, modifying the condition to:
\begin{align}
n_p
+n_{\Sigma^+}
-n_{\Sigma^-}
-n_{\Xi^-}
-n_e
-n_\mu
+n_{K^+}
-n_{K^-}
=0\,.
\end{align}

\subsection{$\beta$ equilibrium}

Neutron stars are expected to reach $\beta$ equilibrium in about a minute after the supernova explosion that gives rise to them. Without kaon condensation, $\beta$ equilibrium involving the entire baryon octet, electrons, and muons reads (from \Cref{mu})
\begin{align}
&\mu_p = \mu_B + \mu_Q\,,
 \nonumber\\
&\mu_n= \mu_B\,,
 \nonumber\\
&\mu_\Lambda = \mu_B\,,
 \nonumber\\
&\mu_\Sigma^+ = \mu_B + \mu_Q\,,
 \nonumber\\
&\mu_\Sigma^0 =\mu_B\,,
 \nonumber
 \end{align}
 \begin{align}
&\mu_\Sigma^- = \mu_B -\mu_Q\,,
 \nonumber\\
&\mu_\Xi^0 = \mu_B\,,
 \nonumber\\
&\mu_\Xi^- =\mu_B - \mu_Q\,,
 \nonumber\\
&\mu_e = - \mu_Q\,,
 \nonumber\\
&\mu_\mu = - \mu_Q\,.
\end{align}

When kaons condense, their presence becomes part of $\beta$ equilibrium
\begin{align}
&\mu_{K^+}=\mu_Q=-\mu_e\,,
\nonumber\\
&\mu_{K^-}=-\mu_Q=\mu_e\,,
\nonumber\\
&\mu_{K^0}=0\,,
\nonumber\\
&\mu_{\bar K^0}=0\,.
\label{kk}
\end{align}
The appearance of additional degrees of freedom modifies the charged chemical potential, significantly affecting the composition of dense matter.

\subsection{Threshold}

S-wave condensation takes place when the energy of the kaon \Cref{eq:disp} falls below its chemical potential
\begin{align}  
\omega_{K}(\vec{k}=0)=m^*_{K}<\mu_{K}\,,
\end{align}
which for the different kaons reads (using \Cref{kk})
\begin{align}  
&m^*_{K^+}<-\mu_e\,,\nonumber
\\ 
&m^*_{K^-}<\mu_e\,,\nonumber
\\
&m^*_{K^0}<0\,,\nonumber
\\
&m^*_{\bar{K}^0}<0\,.
\end{align}

The onset of kaon condensation softens the equation of state, exchanging fermions for condensed kaons, reducing the pressure for a given energy density, which has significant implications for neutron star masses and radii. The presence of kaons can thus play a critical role in determining the stability, dynamics, and observable properties of compact astrophysical objects.

\subsection{Optical potential}

There is substantially less data available to constrain matter that contains strange particles when compared to the non-strange case. One of the ways to verify if dense matter with kaons can connect to nuclear experiments involving kaons is to calculate the optical potential for isospin-symmetric matter at saturation. 

We can write the momentum-dependent optical potential:
\begin{equation}
 U_K=\omega_K-\sqrt{\vec{k}_K^2+m_K^2}\,.
 \end{equation}
When momentum equals to zero, the equation becomes
\begin{equation}
U_K=m_K^*-m_K\,.
 \end{equation} 
The kaon potential is practically the same for the C4 and RC4 couplings. The $\omega\rho$ interaction does not affect isospin-symmetric matter. The potential values we find are $U_{K^-}=-39.27\,\rm{MeV}$ for $d1,\,d2$ and $U_{K^-}=-219.19\,\rm{MeV}$ when using $9\,d1,\,9\,d2$.

The kaon-nucleon interaction is attractive for $K^-$ (not for $K^+$). Theoretical predictions for $K^-$ quote values $U=-140\rm{\,to}-40\,\rm{MeV}$~\cite{Ramos:1999ku}, while experiments provide values around $U=-180\rm{\,to}-200\,\rm{MeV}$~\cite{Friedman:1994hx,Friedman:1998xa}. See Ref.~\cite{Thakur:2025axg} for a recent discussion of stellar properties in the context of kaon potential. Also note that our results (concerning stellar masses) are much less sensitive to the kaon potential than early kaon condensation works~\cite{Glendenning:1997ak}.

%%%%%%%%Figure 1%%%%%%%%%
\begin{figure*}[t]
    \centering
    \includegraphics[width=0.485\textwidth,trim=0 0 0 0,clip]{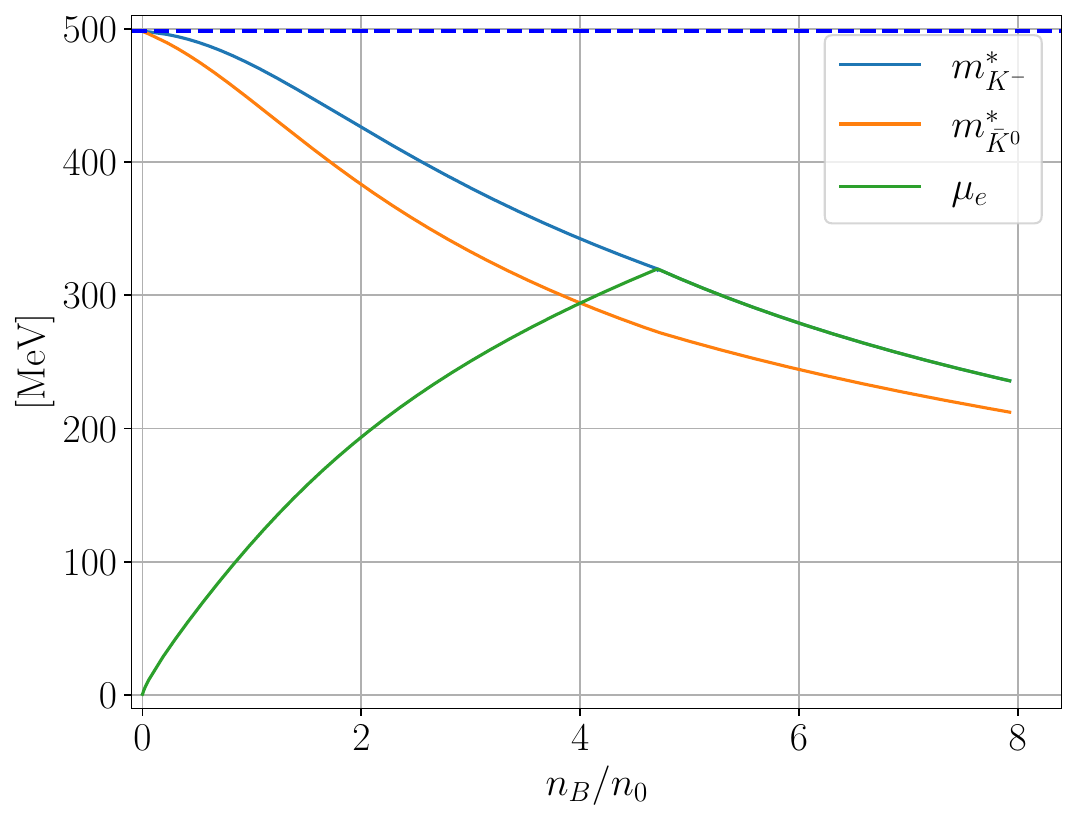}
    \hfill
    \includegraphics[width=0.49\textwidth,trim=0.11cm 0 0 0,clip]{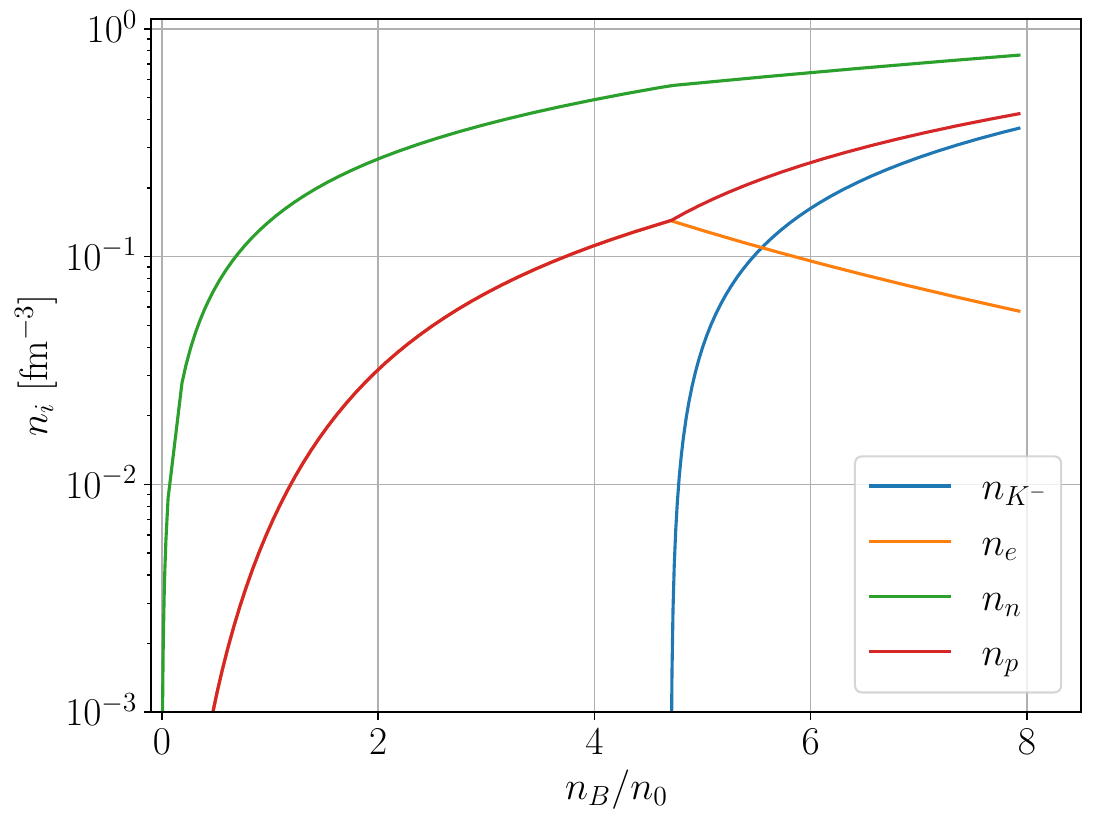}
    \caption{C4 coupling with nucleons, electrons, and kaons. Left: in-medium kaon mass and electron chemical potential as functions of baryon density. Right: particle number density as a function of baryon density. The dashed line marks the kaon vacuum mass.}
    \label{fig:nucleons}
\end{figure*}

\subsection{Energy density}

Although the pressure of matter is modified by the appearance of kaons within our formalism, it is only indirectly, as the condensed kaons do not contribute to the pressure. The indirect effect is carried through changes in $\mu_Q$ and meson-mean fields after the condensation. These affect the pressure not only through to the different particle composition but also due to self-interactions of meson mean and explicit symmetry-breaking terms, including the feedback we previously discussed.

For the energy density, on the other hand, in addition to the indirect effect discussed above, there is a direct contribution from the condensed kaons, which at zero temperature reads
\begin{equation}
\varepsilon_{K}= \mu_{K} n_{K}\,.
\end{equation}

\subsection{Cooling}

The thermal evolution of neutron stars is governed by the general relativistic equations of thermal energy balance and heat transport
\cite{Page:2005fq,Weber:1999qn,Schaab:1996jm}
\begin{align}
  \frac{ \partial (l e^{2\phi})}{\partial m}& = 
  -\frac{1}{\mathcal{E} \sqrt{1 - 2m/r}} \left( \epsilon_\nu 
    e^{2\phi} + c_v \frac{\partial (T e^\phi) }{\partial t} \right) \, , 
  \label{coeq1}  \\
  \frac{\partial (T e^\phi)}{\partial m} &= - 
  \frac{(l e^{\phi})}{16 \pi^2 r^4 \kappa \mathcal{E} \sqrt{1 - 2m/r}} 
  \label{coeq2} 
  \, .
\end{align}

To solve \Cref{coeq1,coeq2}, it is necessary to incorporate both microscopic and macroscopic information. The microscopic description determines quantities associated with the EoS, such as the particle densities, momenta, and effective masses. These quantities are then used to evaluate the specific heat, thermal conductivity, and neutrino luminosity. On the macroscopic side, solving the TOV equations provides the global stellar structure, including the mass, radius, and spacetime curvature, as well as the radial distribution of the microscopic properties mentioned above. Appropriate boundary conditions are also required, and include the condition that the luminosity vanishes at the stellar center and constraints determined by the thermal properties of the stellar atmosphere \cite{1982ApJ...259L..19G,Gudmundsson1983,Page:2005fq}.
In this work, we adopt the simplest atmospheric model, assuming the absence of strong magnetic fields and neglecting any accreted material from fallback.

%%%%%%%%%%%%%%% Figure 2 %%%%%%%%%%%%%%%
\begin{figure*}[t]
    \centering
    \includegraphics[width=0.485\textwidth]{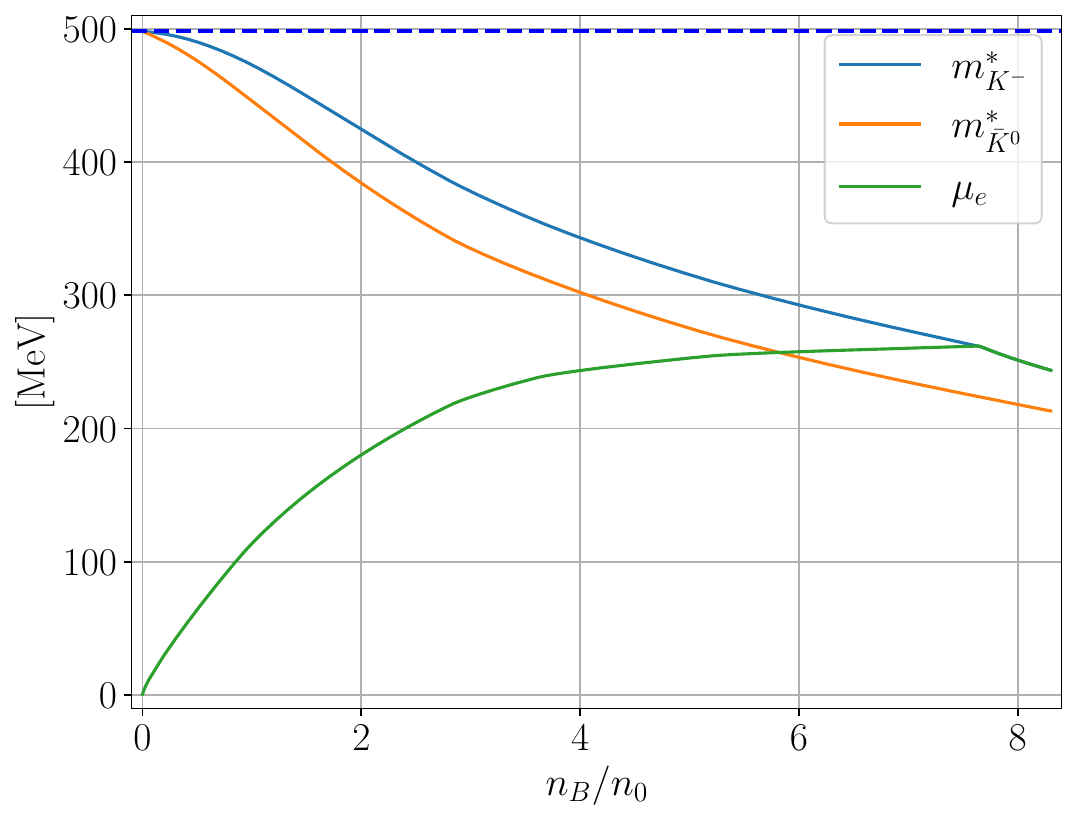}
    \includegraphics[width=0.49\textwidth, trim=0.11cm 0 0 .0cm, clip]{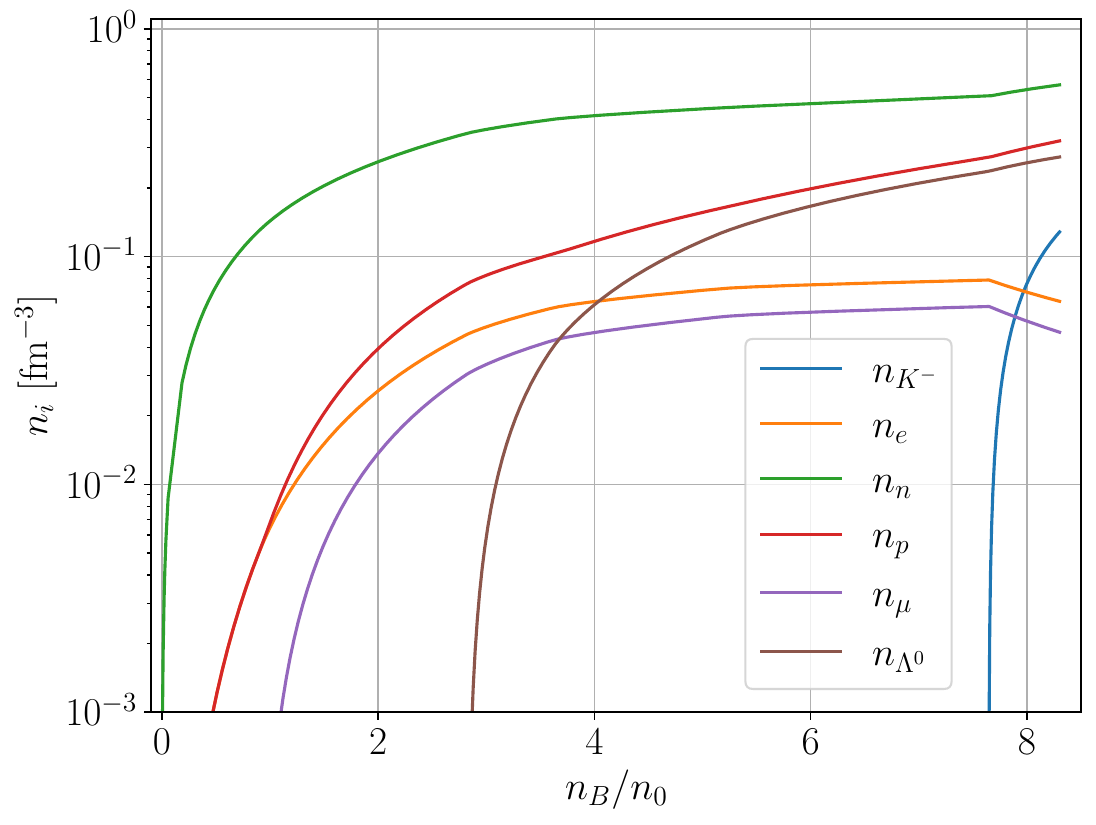}
    \caption{Same as \Cref{fig:nucleons} but including hyperons and muons.}
    \label{fig:hyperons}
\end{figure*}

We then obtain numerical solutions of~\Cref{coeq1,coeq2} for neutron stars spanning a wide range of stellar masses, as well as without and with kaon condensation, hyperons, and muons, and assuming different parametrizations and couplings. In these calculations, we include all relevant neutrino emission mechanisms. The most important among them are the direct Urca (DU), modified Urca (MU), and nucleon–nucleon Bremsstrahlung (BR) processes, which occur in the stellar core. For comprehensive discussions of these mechanisms, refer to \cite{Yakovlev:2000jp,Yakovlev:2004iq}.

The DU process warrants particular attention, as it is significantly more efficient than other neutrino emission mechanisms. The DU process involves neutron beta decay, $n \rightarrow 
p ~+~ e^{-} + \bar{\nu} $, 
, and the inverse reaction of proton electron capture, $p~ +~  e^{-} \rightarrow n + \nu$. 
This mechanism produces an extremely large neutrino luminosity, on the order of $\sim 10^{27} (T_9)^6\,\rm{erg/cm}^3\rm{s}$ (where $T_9$ is short for $T\rm{(in\,K)}/10^9$).
However, it can only occur when the momentum conservation condition $k_{fn} \leqslant k_{fp} + k_{fe}$ is fulfilled. In practice, this requirement is typically satisfied only when the proton fraction exceeds threshold of roughly $11-15\%$~\cite{Lattimer:1991ib,Page:2005fq}. This aspect is particularly relevant for this work because kaon condensation and hyperons can significantly affect the proton Fermi momentum, thereby altering the proton fraction in neutron star matter. 

In addition, the kaon condensate affects the
the nucleon states. transforming them into quasi-particle (mixed) states, which are the coherent superpositions of nucleon states and hyperon-like excitations. The main neutrino reaction associated with these states is again the DU process.
The emissivity for this kaon induced DU,  $\rm{DU}_{\rm{kaon}}$, is associated with the angle $\theta$ described in \Cref{{2D}} and it is orders of magnitude lower than the standard DU process. See Ref.~\cite{Yakovlev:2000jp} for more details.

Combining these two effects, the different EoSs we discuss in the following are expected to produce notable changes in the thermal evolution of neutron stars, even in  cases for which the global macroscopic properties of stars, such as mass and radius, remain approximately the same.

\section{Results}
\label{results}

\subsection{Microscopic properties and stellar structure}
\label{Microscopic properties and stellar structure}

%%%%%%%%%%%%%%% Figure 3 %%%%%%%%%%%%%%%
\begin{figure*}[t]
    \centering
    \includegraphics[width=0.49\textwidth]{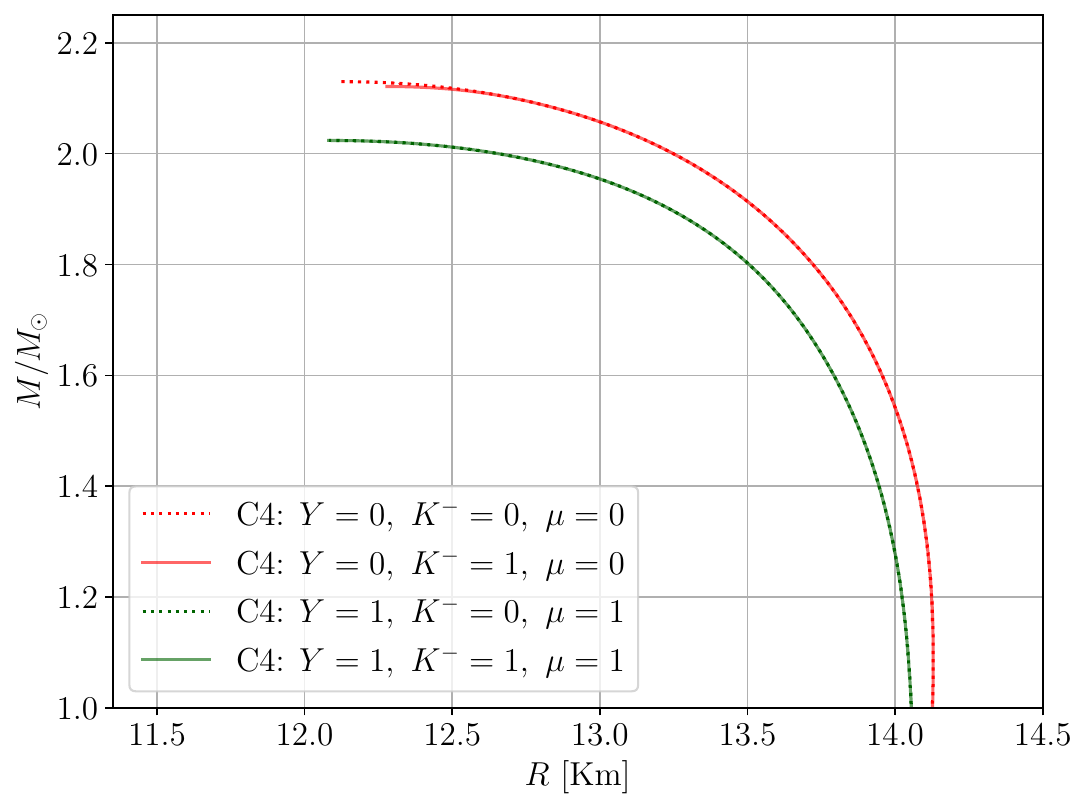}
    \includegraphics[width=0.49\textwidth]{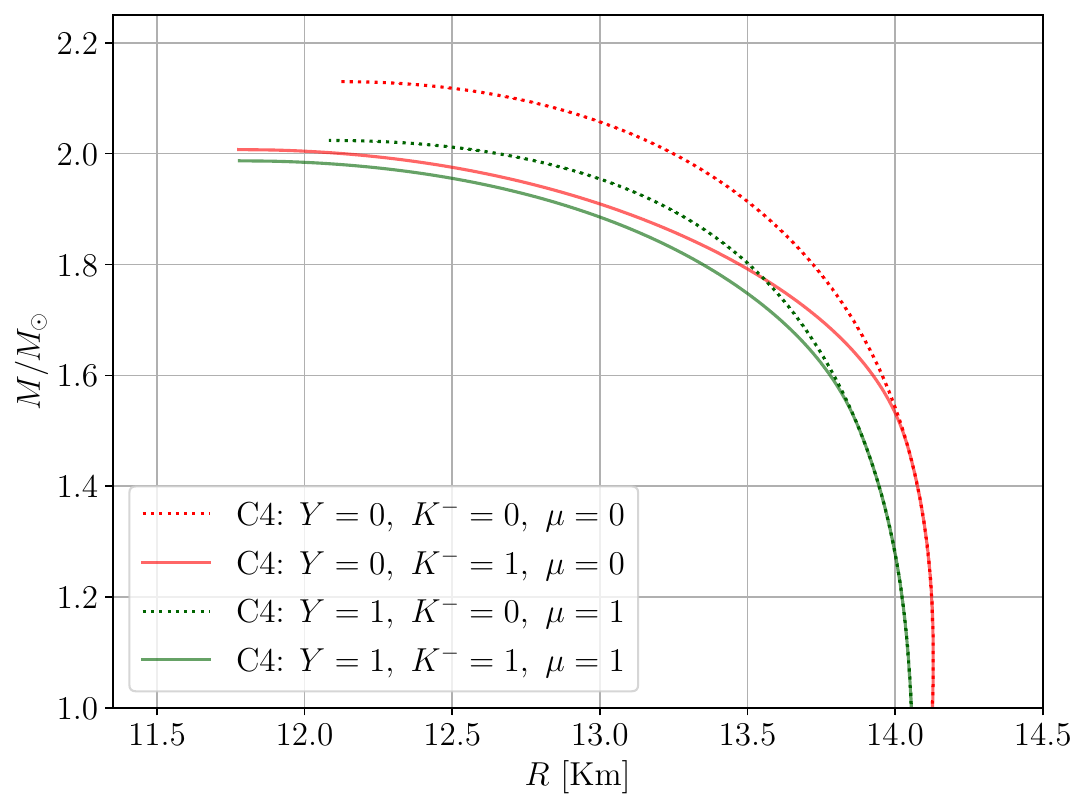}
    \includegraphics[width=0.49\textwidth]{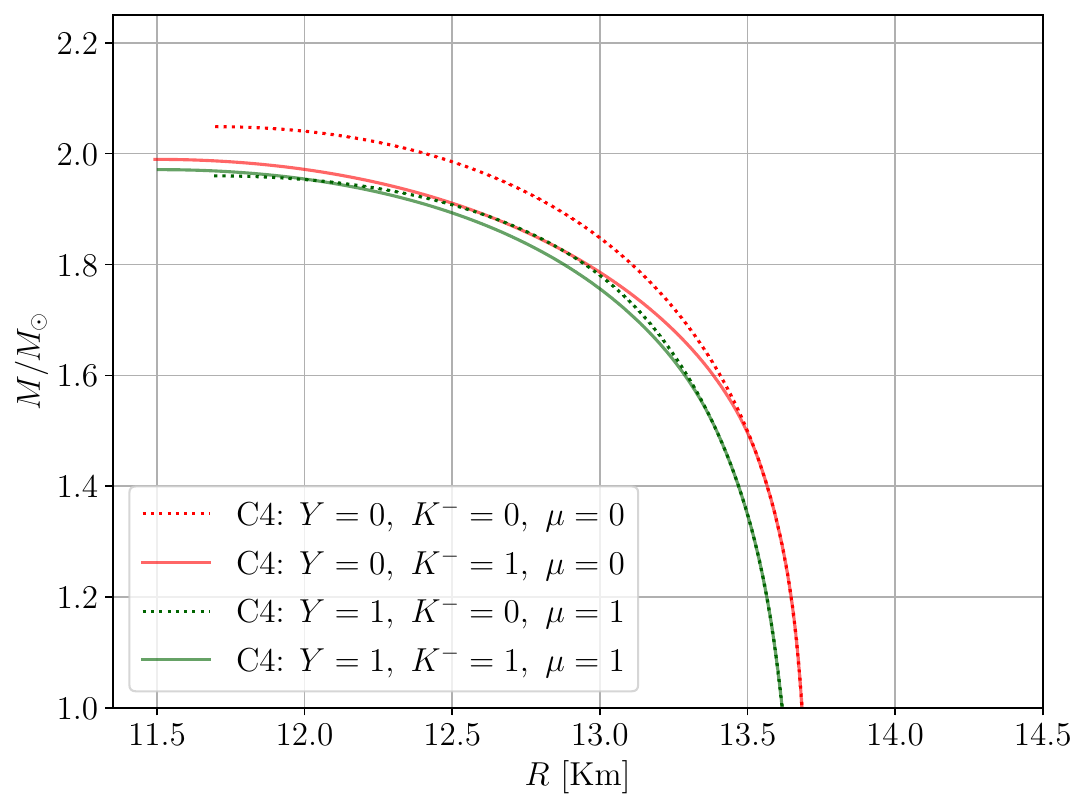}
    \caption{Mass-radius diagrams for C4 coupling with nucleons and electrons, without (0)/with (1) hyperons (Y), without (0)/with (1) kaons ($K^-$), and without (0)/with (1) muons ($\mu$). Top left: original kaon couplings $d_1$ and $d_2$. Top right: $9\,d_1,\,9\,d_2$. Bottom: $9\,d_1,9\,_2$ with additional $\omega\rho$ interaction.}
    \label{fig:MassRadius}
\end{figure*}

%%%%%%%%%%%% Explain Fig 1 %%%%%%%%%%%%
Unlike fermions, bosons do not exist at zero temperature. An exception is the case in which the system lowers its energy by creating a large number of bosons in the ground state (zero momentum), giving rise to a macroscopic boson field. In the case of kaon couplings $d1$ and $d2$ within the C4 parametrization, this takes place for nucleonic neutron-star matter (with electrons) described by mCMF at $n_B=4.7\,n_0$, where the effective mass of $K^-$ meets its effective chemical potential; see left panel of \Cref{fig:nucleons}. For $\bar{K}^0$ this only takes place when its effective mass crosses zero, at asymptotically large densities. For the other two mesons, $K^+$ and $K^0$ their masses remain around the vacuum value in all cases we study, so these mesons never condense. For this reason, we don't show $K^+$ and $K^0$ in the figures.

%%%%%%%%%%%% Explain Fig 1 %%%%%%%%%%%%
As shown in the right panel of \Cref{fig:nucleons}, crossing the $\mu_e^*$ threshold immediately triggers the production of $K^-$, which take the place of most of the electrons balancing the electric charge of protons. The exact balance is not only based on charge and mass, but also on the hadronic interactions, as described in the previous section.

%%%%%%%%%%%%%%% Figure 4 %%%%%%%%%%%%%%%
\begin{figure*}[t]
    \centering
    \includegraphics[width=0.485\textwidth, trim= 0 0 0 0, clip]{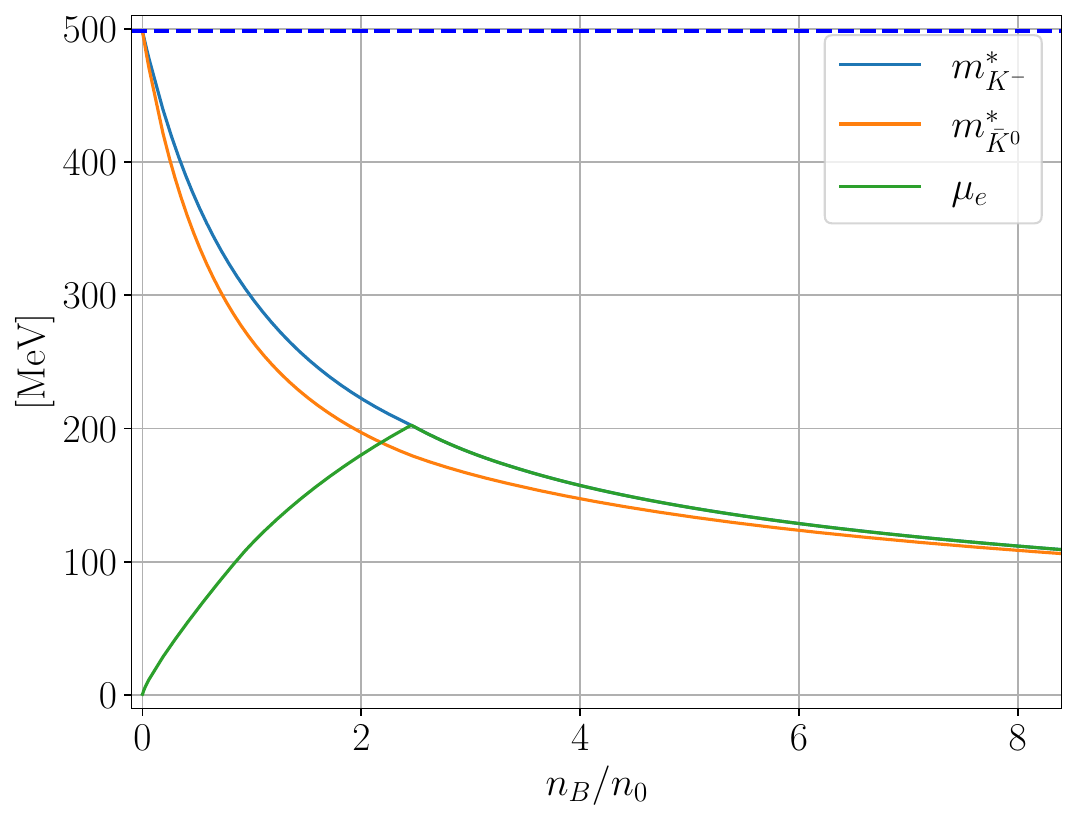}
    \includegraphics[width=0.49\textwidth, trim= 0.11cm 0 0 0, clip]{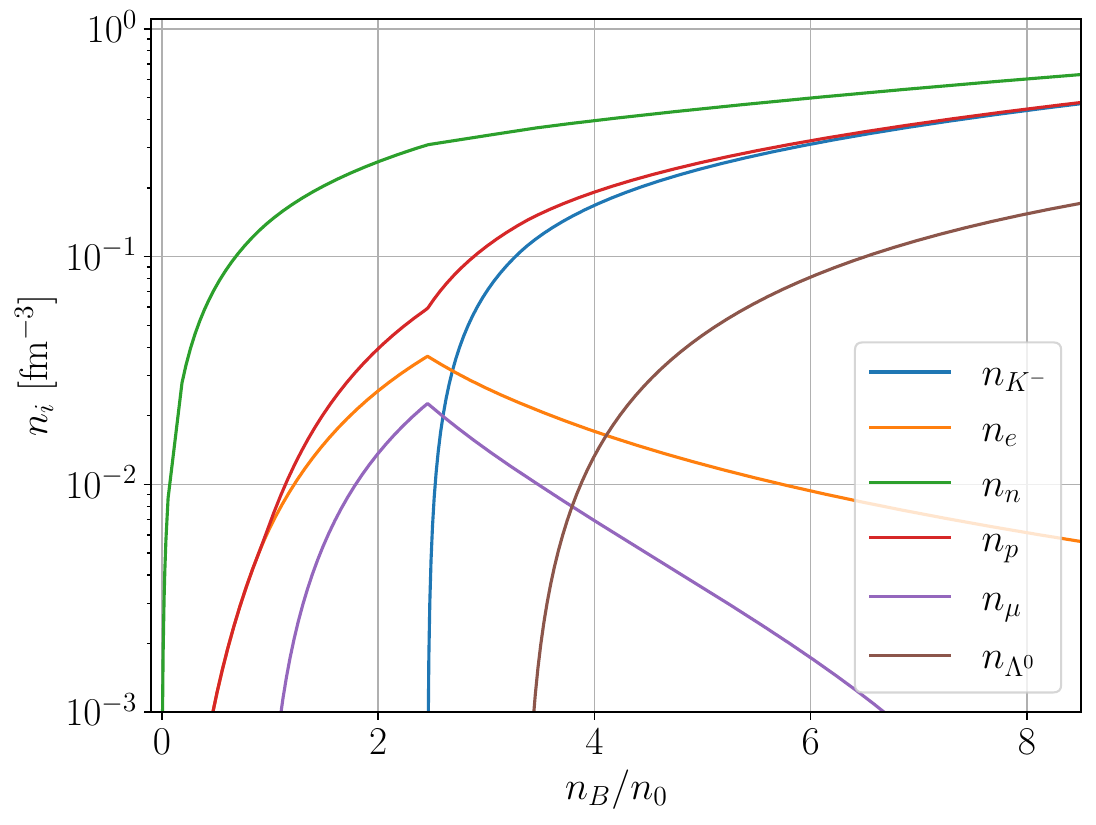}
    \includegraphics[width=0.485\textwidth, trim=0 0cm 0 0, clip]{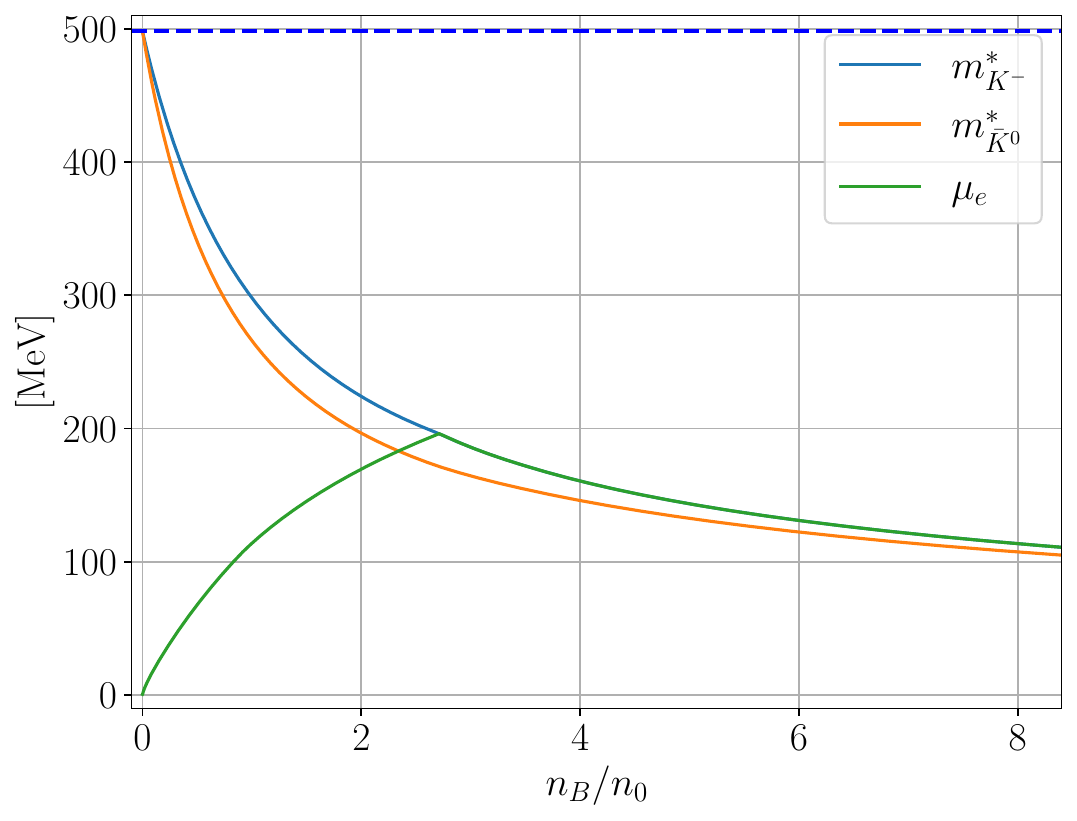}
    \includegraphics[width=0.49\textwidth, trim=0.11cm 0 0 0, clip]{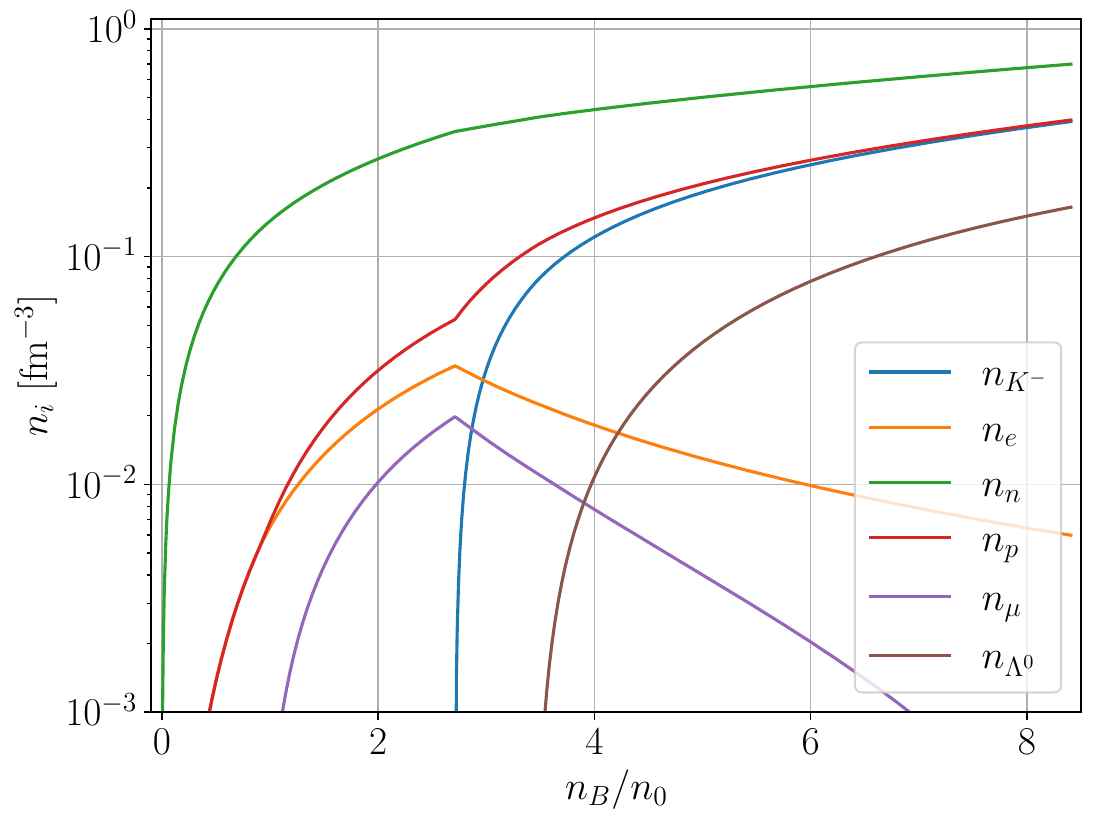}
    \caption{Same as \Cref{fig:hyperons} but with $9\,d_1,\,9\,d_2$ (top panels) and $9\,d_1,\,9\,d_2$ plus additional $\omega\rho$ interaction (bottom panels).}
    \label{fig:9times_omegarho}
\end{figure*}

%%%%%%%%%%%% Explain Fig 2 %%%%%%%%%%%%
Once the hyperons are included in the calculation, the picture is qualitatively the same, but quantitatively the condensation is shifted to larger densities. As shown in the left panel of \Cref{fig:hyperons}, for neutron star matter with nucleons and hyperons (and electrons and muons) described by mCMF, the effective mass of $K^-$ meets its effective chemical potential at $n_B=7.6\,n_0$. The $\bar{K}^0$'s effective mass crosses zero once more far beyond the scale of our figure. The right panel of \Cref{fig:hyperons} shows that crossing the $\mu_e^*$ threshold immediately triggers the production of $K^-$, which take place of the most of the electrons and muons balancing the electric charge of protons at high densities. At this point, the $\Lambda$ hyperons have already reached densities comparable to the protons. The other hyperons only appear at even larger densities within the C4 parametrization. This result agrees with other calculations that show that hyperons suppress kaon condensation in neutron stars (see e.g., Ref.~\cite{Mishra:2009bp} with implications for protoneutron stars). This happens because the hyperons shift $\mu_e^*$ to lower values once they appear.

%%%%%%%%%%%% Explain Fig 3 %%%%%%%%%%%%
In the top left panel of \Cref{fig:MassRadius} we present the mass-radius diagram for the cases we discussed, distinguishing between families (stellar sequences calculated assuming one EoS but different stellar central densities) that do include or do not include kaon condensation, hyperons, and muons. Although we don't focus on low-mass stars, we have added a Baym, Pethick and Sutherland
(known as BPS) crust, which includes an inner crust, an
outer crust, and an atmosphere~\cite{osti_4718088} to our core EoSs. Independently of the consideration of kaon condensation, adding hyperons and muons decreases both the stellar masses and radii (going from red to green curves), mainly mass for the hyperons and radius for the muons. Allowing for kaon condensation (going from dotted to full curves of each color) changes little the mass-radius relation of massive nucleonic stars, beyond $2.1\,M_\odot$, at which the central density is large enough $n_B=4.7\,n_0$ to give rise to $K^-$ condensation. In the case of stars with hyperons (and muons), there is no difference because the maximum mass has a density of $n_B=6.1\,n_0$, bellow the threshold for condensation, $n_B=7.6\,n_0$.

%%%%%%%%%%%% Explain Fig 3 %%%%%%%%%%%%
To understand the effect of the strength of kaon couplings, $d_1$ and $d_2$, we explore the effects of larger couplings, $9\times$ the original ones. This corresponds to using scattering lengths for the different channels of $a(I=0)=2.09$ and $a(I=1)=3.82$.
Our results for $9\,d_1$ and $9\,d_2$ are  shown in the top right panel of \Cref{fig:MassRadius}. Going beyond $9\times$ creates numerical problems in our formalism. In this case, one can clearly see that the threshold for condensation takes place at much lower star masses, both without and with hyperons and muons ($1.5\,M_\odot$). As before, going from nucleonic stars to stars with hyperons has a pronounced effect in the case without kaons (dotted curves) but, in the case with kaons (full curves), the difference becomes much smaller.

%%%%%%%%%%%%%%% Figure 5 %%%%%%%%%%%%%%%
\begin{figure*}[t!]
    \centering
    \includegraphics[width=0.49\textwidth, trim=0.11cm 0 0 0, clip]{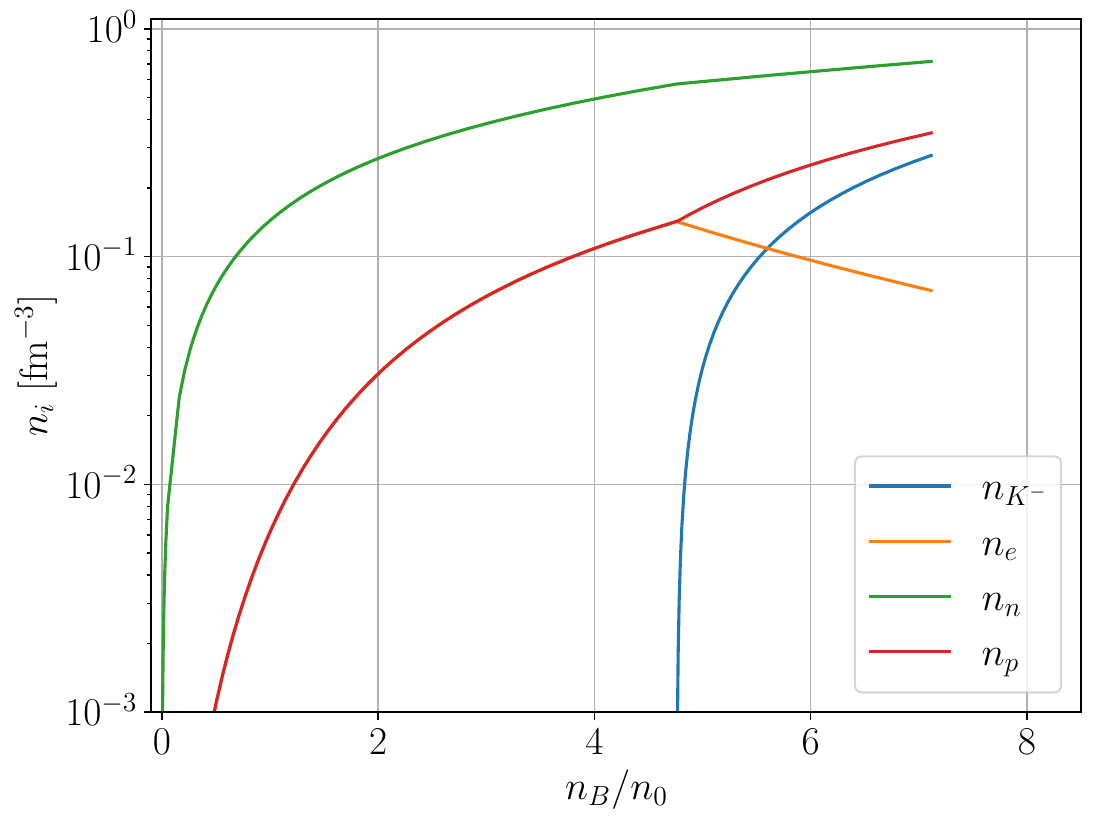}
    \includegraphics[width=0.49\textwidth, trim=0.11cm 0 0 0, clip]{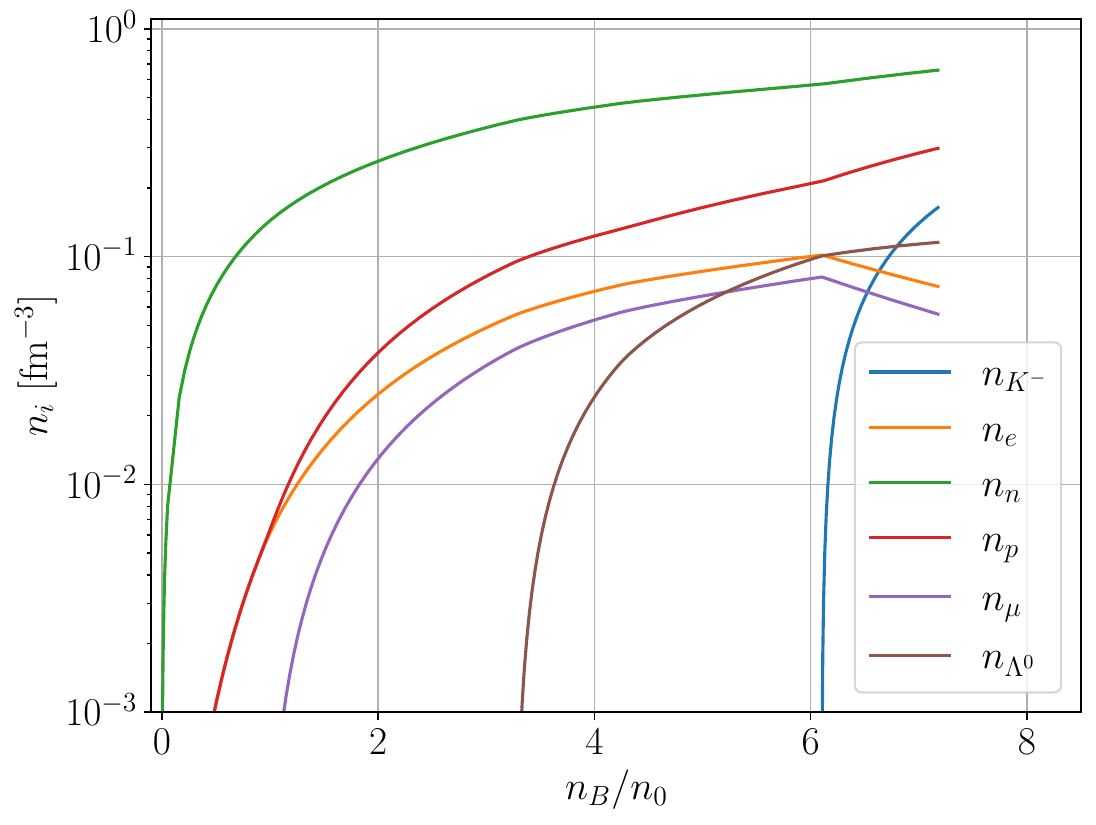}
    \includegraphics[width=0.49\textwidth, trim=0.11cm 0 0 0, clip]{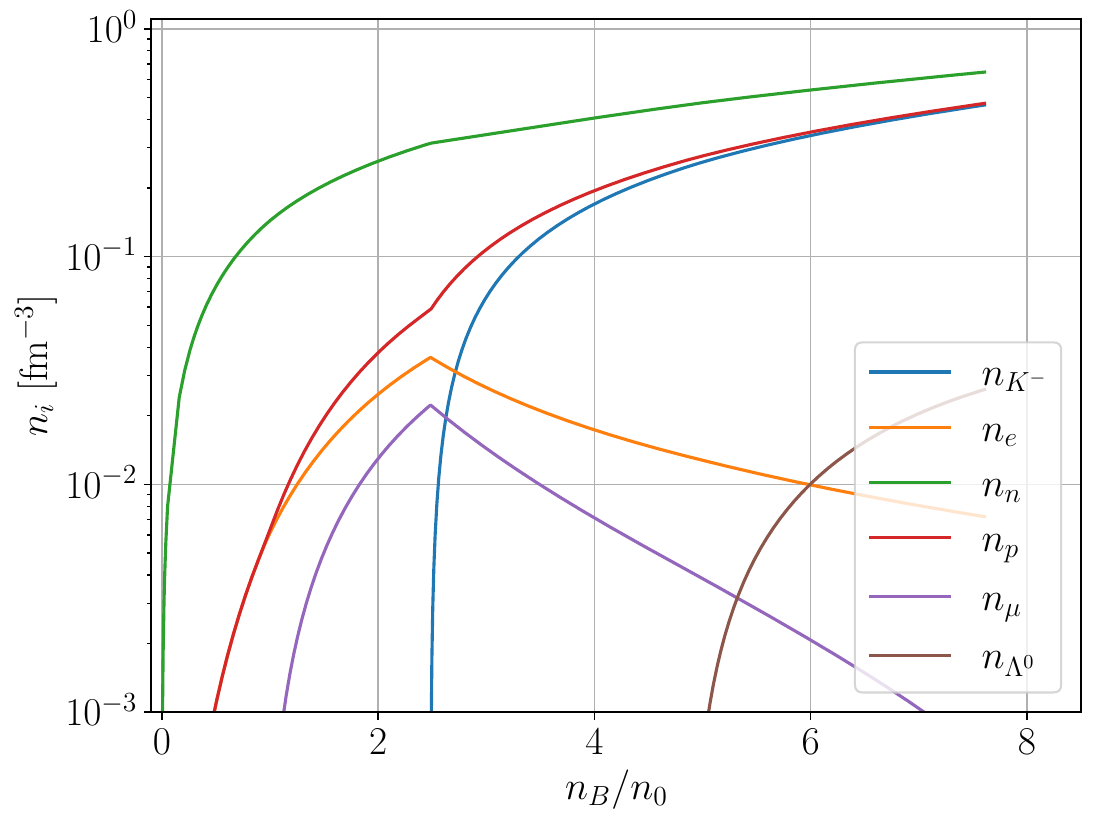}
    \includegraphics[width=0.49\textwidth, trim=0.11cm 0 0 0, clip]{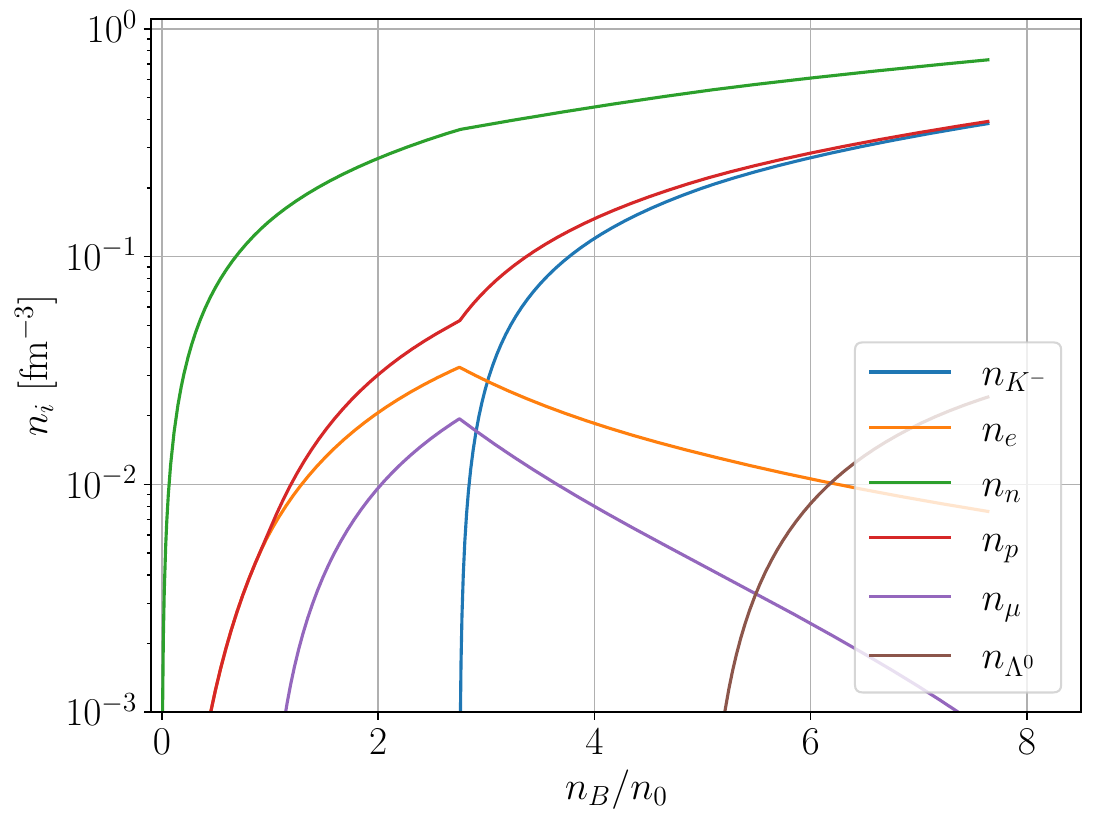}
    \caption{Same as the right panels of \Cref{fig:nucleons}, \Cref{fig:hyperons}, and \Cref{fig:9times_omegarho} showing particle number density, as a function of baryon density but for the RC4 parametrization. Top left: with nucleons, electrons, and kaons. Top right: also including hyperons. Bottom left: also including hyperons with $9\,d_1,\,9\,d_2$. Bottom right: also including hyperons with $9\,d_1,\,9\,d_2$ and additional $\omega\rho$ interaction.}
    \label{fig:RC4nucleonshyperons}
\end{figure*}

%%%%%%%%%%%% Explain Fig 4 %%%%%%%%%%%%
As seen in the top panels of \Cref{fig:9times_omegarho}, the threshold for $K^-$ condensation in the presence of hyperons (and muons) for $9\times$ $d_1$ and $d_2$ is $n_B=2.5\,n_0$, before the appearance of the first hyperons, the $\Lambda$, at $n_B=3.3\,n_0$. This is a consequence of the sharp decrease in the effective mass of the kaons that takes place when the kaons appear. In this case, there is a softening in the EoS due to the kaon condensation, as well as later for the hyperons, and the hyperons do not suppress the condensation. This is not in agreement with the results previously discussed~\cite{Mishra:2009bp}. In our case, the kaons are the ones suppressing the hyperons, which would otherwise, in the absence of kaons, appear at $n_B=2.8\,n_0$ (not shown here). 
%%%%%%%%%%%% Explain Fig 3 %%%%%%%%%%%%
Interestingly, the effect of kaons and hyperons is similar with respect to the maximum mass of neutron stars (from dotted red curve to dotted green or full red in the top right panel of \Cref{fig:MassRadius}).

%%%%%%%%%%%% Explain Fig 3 %%%%%%%%%%%%
The stellar masses reproduced in the top panels of \Cref{fig:9times_omegarho} agree with observations of neutron stars with masses consistent with $2\,M_\odot$~\cite{Antoniadis:2013pzd,Fonseca:2021wxt}. They are also within the error range provided for neutron star radii based on NICER data for different sources between $10.73\,\rm{km}$~\cite{Choudhury:2024xbk} and $15.01\,\rm{km}$~\cite{Dittmann:2024mbo}, both within one standard deviation. Nevertheless, tidal deformability measured by LIGO for the neutron star merger GW170817~\cite{LIGOScientific:2018hze} $\tilde{\Lambda}<730$ requires smaller stars, although the relation between  $\tilde{\Lambda}$ and stellar radius is not trivial~\cite{Dexheimer:2018dhb}. Such small values of tidal deformability (without decreasing star masses) require different interactions, such as the $\omega\rho$ suggested in Ref.~\cite{Horowitz:2002mb} that we use next. 

%%%%%%%%%%%%%%% Figure 6 %%%%%%%%%%%%%%%
\begin{figure*}[t!]
    \centering
    \includegraphics[width=0.49\textwidth]{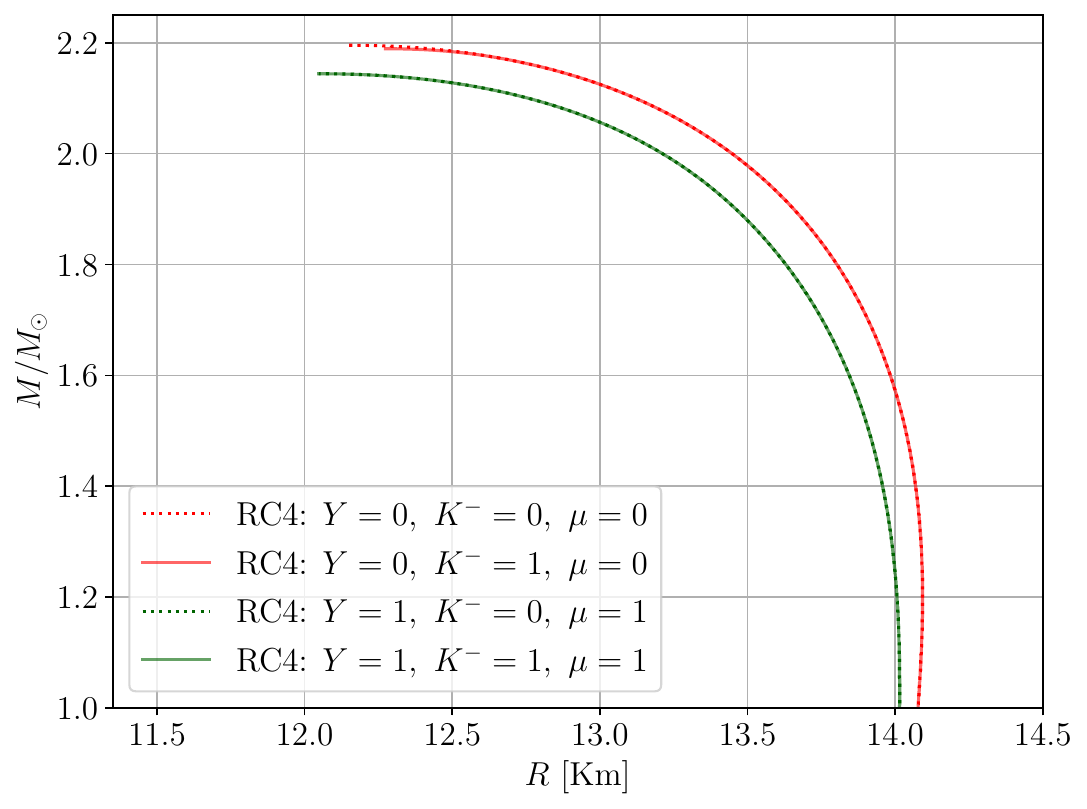}
    \includegraphics[width=0.49\textwidth]{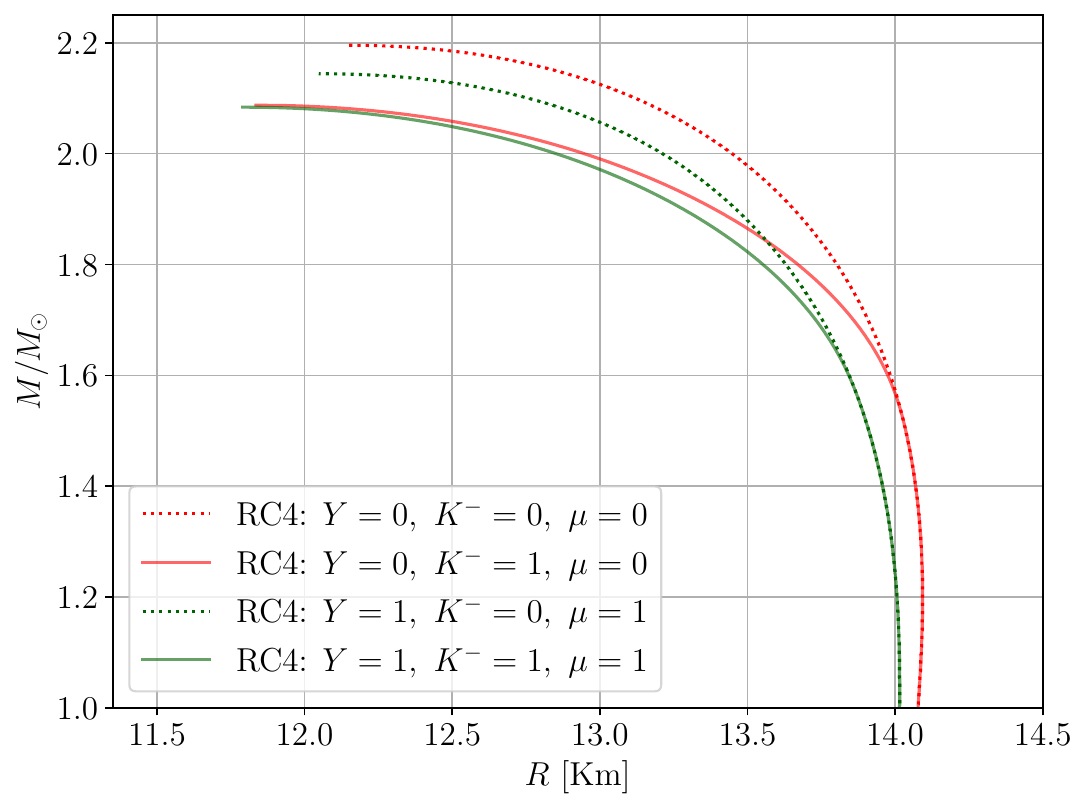}
    \includegraphics[width=0.49\textwidth]{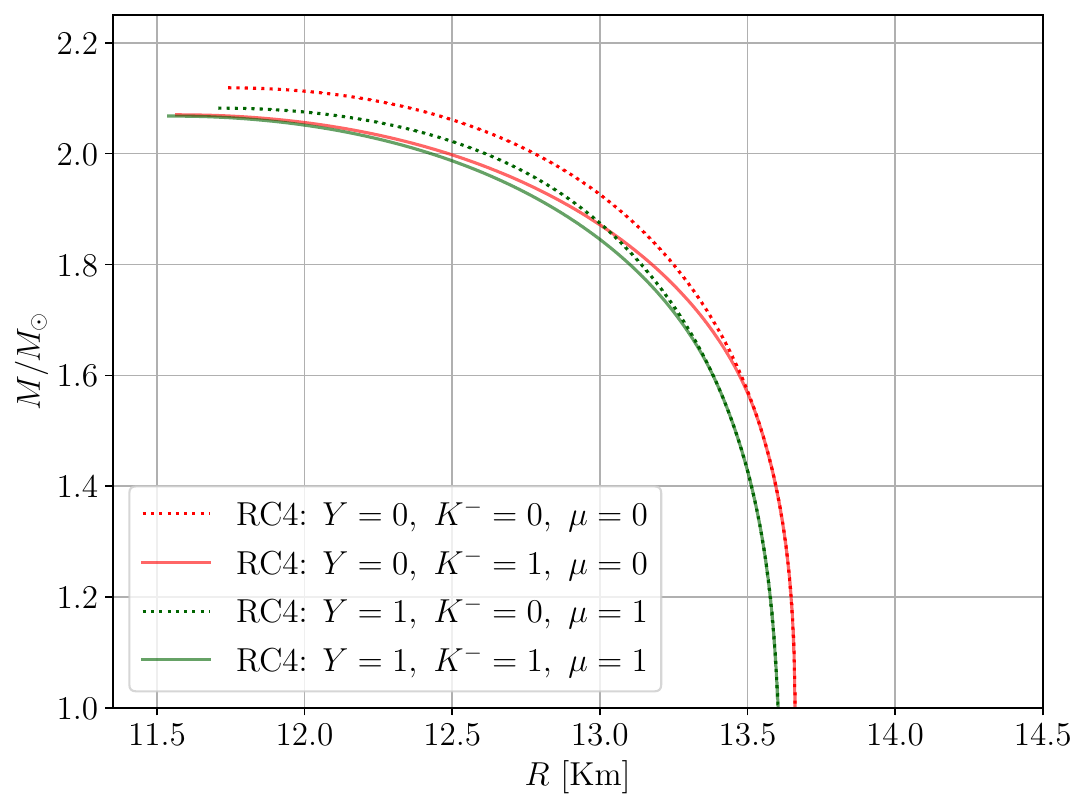}
    \caption{Same as \Cref{fig:MassRadius} and but for the RC4 parametrization.}
    \label{fig:RC4MassRadius}
\end{figure*}

%%%%%%%%%%%% Explain Fig 3 %%%%%%%%%%%%
In the CMF C4 coupling, the $\omega\rho$ interaction is calibrated to reproduce the tidal deformability measured by LIGO. The results with the additional $\omega\rho$ interaction are shown in the lower panel of \Cref{fig:MassRadius} for the case with $9\,d_1$ and $9\,d_2$.  While all stars present a smaller radius by about $1/2$ of a km (when compared to the top right panel), they also present slightly lower stellar masses, more for the cases without kaon condensation. In one case (with hyperons - green curves), considering kaon condensation actually increases the possible neutron star masses. 

%%%%%%%%%%%% Explain Fig 4 %%%%%%%%%%%%
To understand why that is the case, we look at the bottom panels of \Cref{fig:9times_omegarho}. As in the case of the top panels (without $\omega\rho$ interactions), kaon condensation appears at $n_B=2.5\,n_0$, before the hyperons at $n_B=3.3\,n_0$. This means that the kaons are taking the place of the hyperons, pushing the appearance of hyperons to larger densities and in smaller amounts (than in the case without the kaons, not shown here). When looking at the bottom panels (in comparison with the top ones), the kaons appear at slightly larger densities, there are slightly fewer kaons, as well as protons, because matter is slightly more isospin asymmetric. In this case, kaon condensation takes place at $n_B=2.7\,n_0$ and hyperons appear at $n_B=3.4\,n_0$. These effects from the $\omega\rho$ interaction are in agreement with the conclusions from the RMF works presented in Ref.~\cite{Panda:2013oza,Ma:2022fmu}, but in our case the $\omega\rho$ interaction does not prevent kaon condensation, as in Ref.~\cite{Ma:2022fmu}. Ref.~\cite{Sharifi:2021gvi} also reconciles tidal deformability data with matter with strangeness using the lowest-order constrained variational (LOCV) approach.

%%%%%%%%%%%%%%% Figure 7 %%%%%%%%%%%%%%%
\begin{figure*}[t]
    \centering
    \includegraphics[width=0.49\textwidth, trim=2cm 0 3cm 2cm, clip]
    {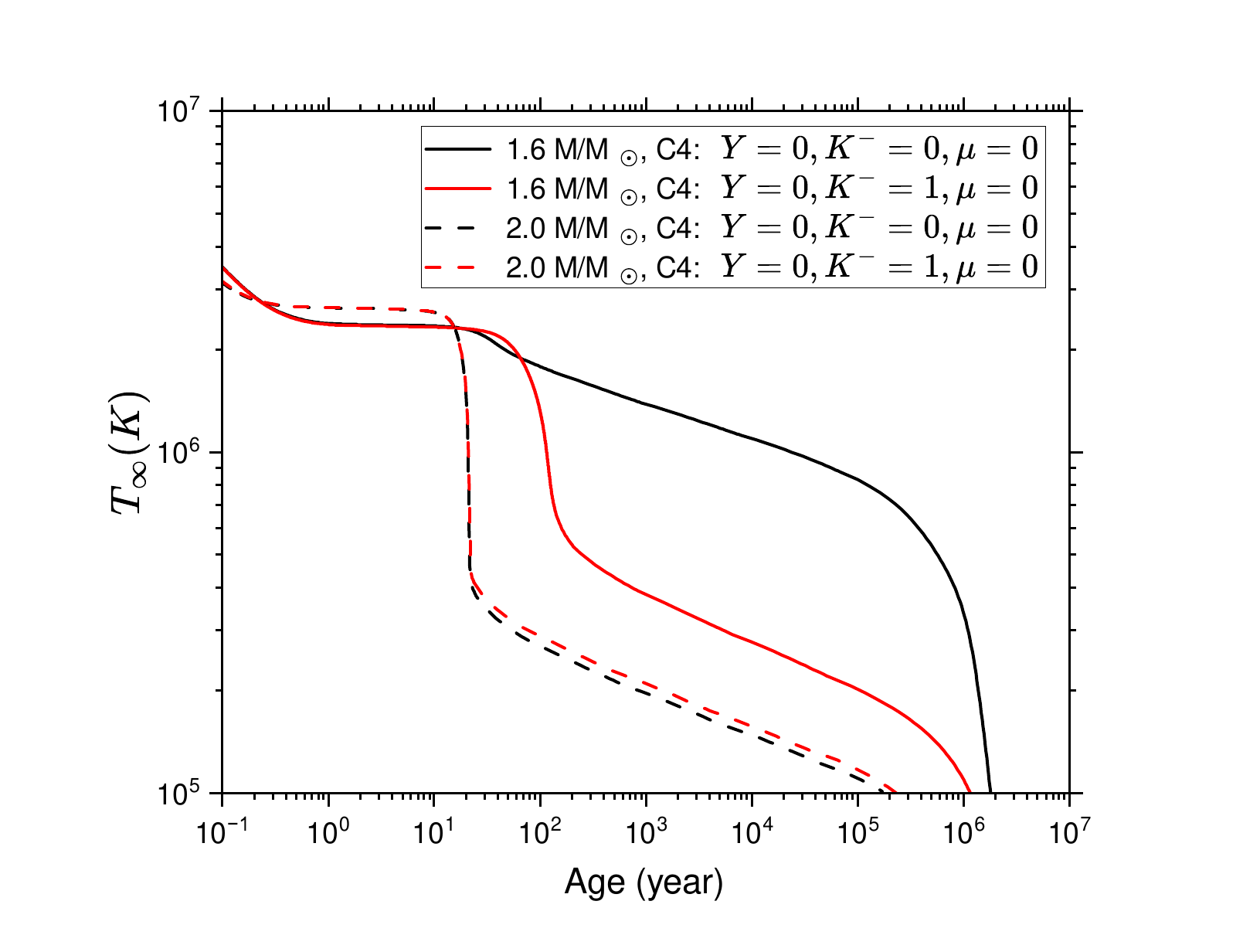}
    \includegraphics[width=0.49\textwidth, trim=2cm 0 3cm 2cm, clip]{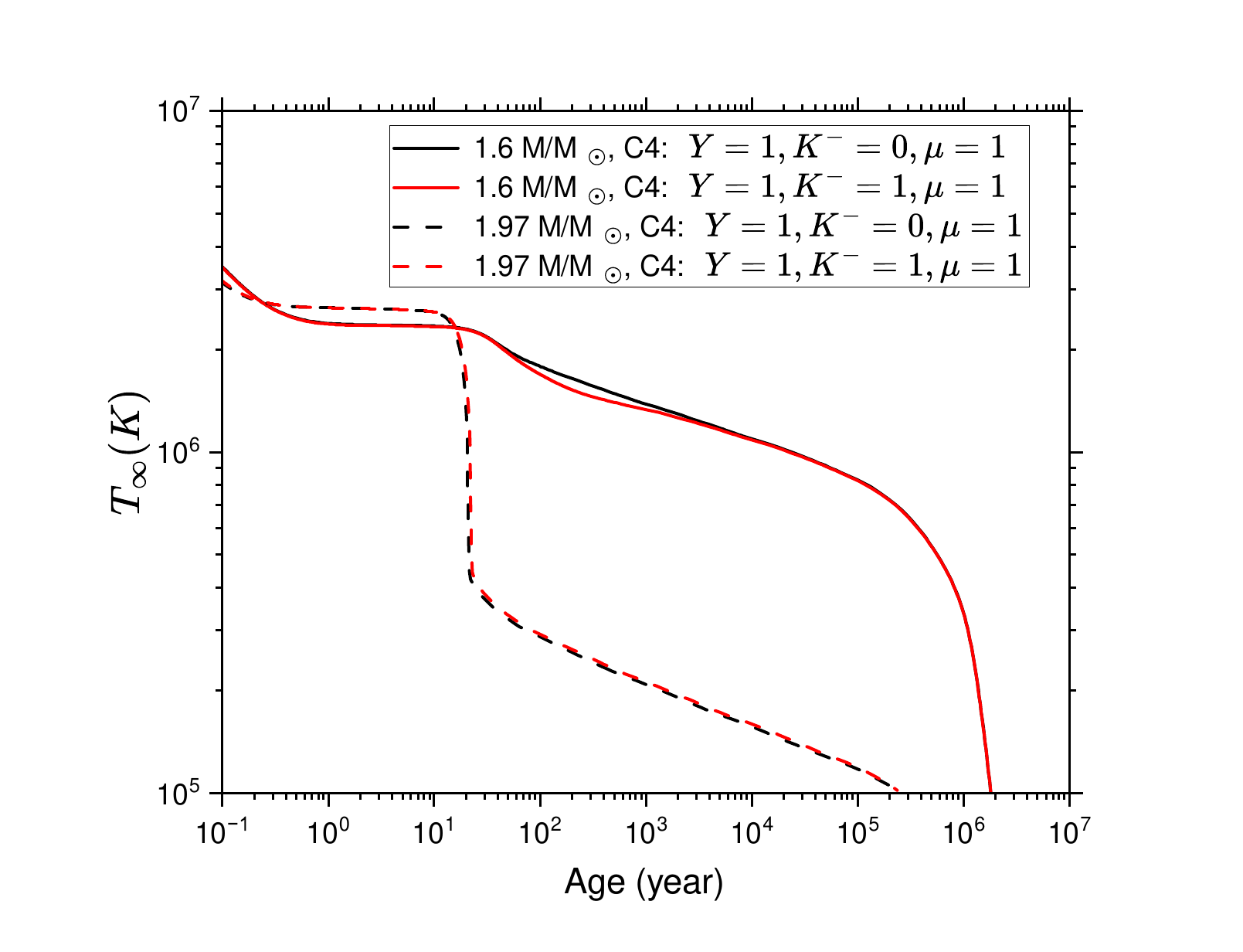}
    \includegraphics[width=0.49\textwidth, trim=2cm 0 3cm 2cm, clip]
    {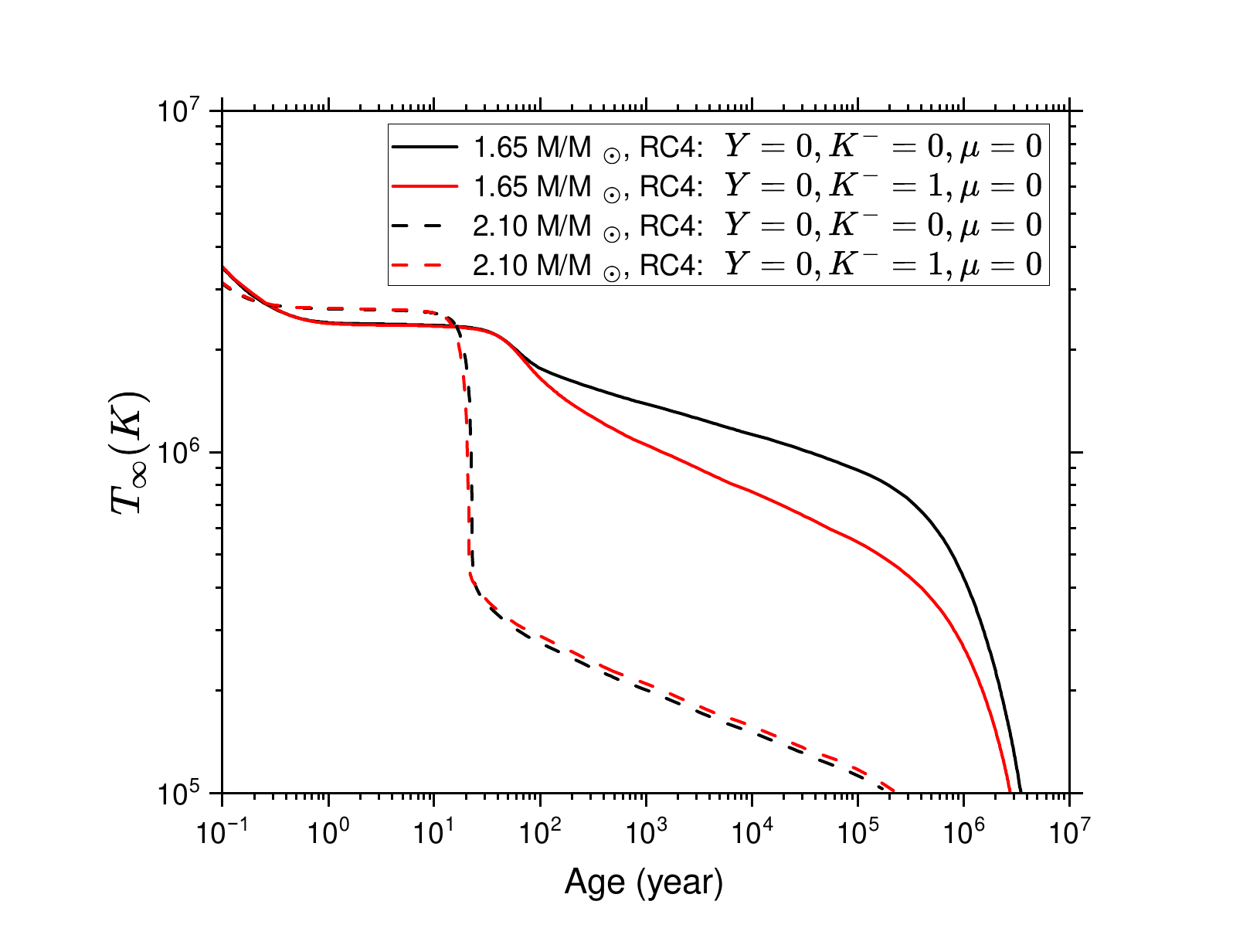}
    \includegraphics[width=0.49\textwidth, trim=2cm 0 3cm 2cm, clip]
    {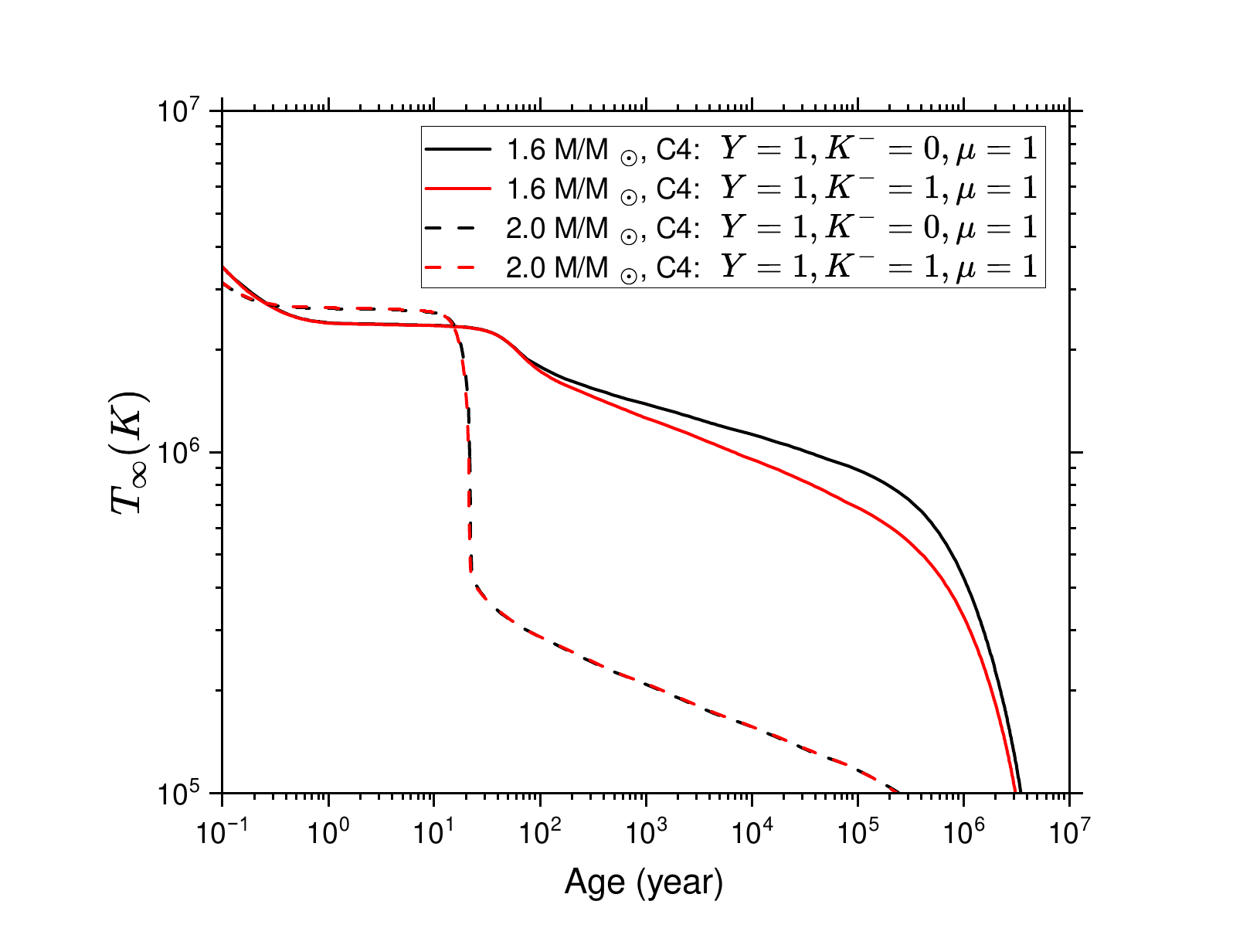}
    \caption{Thermal evolution of neutron stars showing the temperature, as measured by an observer at infinity, as a function of time after formation. All curves correspond to $9\,d_1,\,9\,d_2$ and additional $\omega\rho$ interaction. Top panels: C4 parametrization. Bottom panels: RC4 parametrization. Left panels: with nucleons and electrons. Right panels: with nucleons, electrons, muons, and hyperons.}
    \label{cooling1}
\end{figure*}

%%%%%%%%%%%% Explain Fig 5 %%%%%%%%%%%%
To verify how our results depend on a specific parametrization, we proceed by redoing our analysis for another parametrization of CMF, RC4. The results are shown in \Cref{fig:RC4nucleonshyperons}. First, not including hyperons and muons, the particle population is essentially the same as the one already discussed, with a kaon threshold of $n_B=4.8\,n_0$ (top left panel). Next, with hyperons and muons, there are fewer hyperons now, which also appear at larger densities, but the kaons appear significantly earlier, at $n_B=6.1\,n_0$ (top right panel). For $9\,d_1$ and $9\,d_2$, the difference is that now the hyperons appear much later (bottom left panel - kaon threshold of $n_B=2.5\,n_0$), also when we additionally include  the $\omega\rho$ interaction (bottom right panel - kaon threshold of $n_B=2.8\,n_0$).

%%%%%%%%%%%% Explain Fig 6 %%%%%%%%%%%%
We also calculate mass-radius relations for a family of neutron stars within the RC4 parametrization. The results are presented in \Cref{fig:RC4MassRadius}. Overall, when comparing with \Cref{fig:MassRadius}, RC4 presents larger neutron star masses in all cases, and the (green) curves with hyperons and muons don't present a large mass reduction with respect to the respective red ones. More importantly, the $\omega\rho$ interaction included in the lower panel does not have such a strong effect as for C4, although it also brings several curves to present the same maximum stellar mass, with hyperons, kaons, or both (all in good agreement with observations).

\subsection{Thermal evolution}
\label{Thermal evolution}

%%%%%%%%%%%% Explain Fig 7 %%%%%%%%%%%%
As previously noted, the neutrino emissivity linked to kaons, $\rm{DU}_{\rm{kaon}}$, is several orders of magnitude weaker than that of the standard Direct Urca process, DU. Consequently, if DU operates in the star, the onset of kaon condensation does not lead to any observable impact on stellar cooling. For this reason, in this section we restrict our attention to scenarios in which kaon condensation occurs (thus enabling the $\rm{DU}_{\rm{kaon}}$ process) but the standard DU process does not operate.
This behavior occurs for $9\,d_1,\,9\,d_2$ with $\omega\rho$. The choice of $9\,d_1,\,9\,d_2$ leads to kaon condensation at comparatively lower densities, whereas the inclusion of $\omega\rho$ shifts the onset of DU to higher densities (so that it appears only in more massive stars), as a consequence of the reduced electron and muon populations. Therefore, we concentrate on intermediate-mass stars, with $M \sim 1.6\,M_\odot$, which represent the lowest masses whose central densities are still sufficient to allow for $\rm{DU}_{\rm{kaon}}$.

%%%%%%%%%%%% Explain Fig 7 %%%%%%%%%%%%
The corresponding curves are displayed in \Cref{cooling1}. Each panel presents the thermal evolution of stars with two distinct masses, comparing models that include condensed $K^-$ (red) with those without condensed $K^-$ (black). As already mentioned, the lower mass, $\sim1.6\,M_\odot$, represents the smallest density at which kaon condensation occurs, and the higher mass, $\sim2\,M_\odot$, represents the larger density reached in stable neutron stars for these EoSs.
We start our discussion with masses $\sim1.6\,M_\odot$.
In the top left panel, there is a substantial difference: stars with kaon condensation cool much more efficiently. When hyperons and muons are included (moving from the top left to the top right panel), this cooling difference is reduced because the presence of muons shifts the onset of $\rm{DU}_{\rm{kaon}}$ to higher densities. When the parametrization is changed from C4 to RC4 (moving from the top to the bottom panels), the impact of kaons on the cooling pattern changes somewhat, but remains qualitatively similar. The main distinction is that, for RC4, kaons emerge at slightly higher densities, so the condensation effect in the bottom left panel is less pronounced than in the top left panel for the same stellar mass. Furthermore, RC4 features fewer hyperons than C4, which makes the difference between the two bottom panels smaller than the corresponding difference between the top panels.
For masses $\sim2\,M_\odot$, the standard DU is already allowed in all cases, so this process dominates and kaon condensation cannot be observed. The small differences between dashed lines of different colors in the bottom left panel of \Cref{cooling1} are due to the fact that DU takes place in a smaller part of the $\sim2\,M_\odot$ stars (as compared to the stars in the right panels with muons and  hyperons), so a tiny effect from kaon condensation can still be observed.

\section{Discussion and conclusions}

Kaon condensation is an important phenomenon expected to take place in dense nuclear matter, particularly in neutron star interiors, where extreme conditions favor the formation of exotic particles. As the density increases, condensed kaons can appear in large amounts, influencing the equation of state and the overall structure of the star. 
Within a chiral formalism at zero temperature, we investigate the interplay between s-wave kaon condensation and the hyperon population in dense matter and, ultimately, the consequences for neutron stars. We make use of a recently updated version of the Chiral Mean Field Model with an improved meson description (mCMF), in which in-medium masses for the pseudoscalar mesons are dynamically calculated from explicit-symmetry-breaking mechanism, allowing them to vanish in the chiral limit, as expected. As a result, our mechanism includes a feedback from the mesons into the equations of motion, in such a way that changes in the in-medium masses of mesons also affect the baryons.

We find that only the $K^-$ condenses within the regime expected to exist in the interior of neutron stars (zero temperature, charge neutral, in $\beta$ equilibrium, densities $<8\,n_0$). While $\bar{K}^0$ almost condenses, the masses of the $K^+$ and $K^0$ remain close to their vacuum values. The density at which the $K^-$ condenses depends on two free parameters, which can be traced to a combination of nucleon-kaon scattering lengths. Since these scattering lengths are still not well constrained, we use a value commonly found in the literature but also explore larger values. We study different particle populations, different CMF couplings, and interactions. As a result, $K^-$ condensation takes place between $n_B=2.5-7.6\,n_0$.

Hadronic CMF parameters have been constrained to reproduce nuclear physics and astrophysics. While C4 is the standard CMF coupling, the new RC4 coupling relies on a renormalization of vector fields that allows for a description of different meson masses without breaking chiral symmetry. While RC4 reproduces larger neutron star masses than C4, all couplings and parametrizations studied in this work reproduce neutron stars (without and with kaon condensation) around $2\,M_\odot$ and radii in agreement with NICER results. The addition of the $\omega\rho$ interaction further ensures that tidal deformability constraints obtained from gravitational waves are reproduced (see Ref.~\cite{Parmar:2024qff} for a study of Bayesian inference of kaon condensation and neutron star properties within a RMF model).

Our main finding is that, in disagreement with works in the literature, the appearance of hyperons does not necessarily suppress kaon condensation. Using the same parameters as other works, $K^-$ condensation takes place in high-mass neutron stars, although it does not significantly modify mass-radius curves. Nevertheless, using different parameters, $K^-$ condensation takes place inside lower-mass stars, and depending on the parameters, kaons appear at lower densities than hyperons, effectively suppressing them. Specifically, the $\omega\rho$ interaction increases isospin asymmetry while decreasing the electron and muon chemical potentials, thereby decreasing the density of both hyperons and negative kaons. Although this effect also decreases the maximum mass allowed for neutron stars, this is not the case (significantly) in the presence of kaon condensation, where for C4 the condensation of $K^-$ actually increases the allowed stellar mass. 

Having established that kaon condensation can appear in neutron stars, even in competition with hyperons, how can it be observed, given that it can produce effects on the stellar mass–radius relation similar to those of hyperons? The thermal evolution of neutron stars can help answering this question. In particular, for stellar masses that reproduce stellar central densities allowing for a kaon-induced direct Urca process,  $\rm{DU}_{\rm{kaon}}$, kaon condensation makes these stars cool notably faster. Nevertheless, this behavior can only be observed for certain parametrizations for intermediate- to high-mass stars that allow for kaon condensation and couplings that do not allow for the standard DU process in these stars. We find the range of masses to be $\sim1.6-~2\,M_\odot$  and for the parametrizations that reproduce kaon condensation at lower densities and include the $\omega\rho$ interaction. 

While our work allows for the inclusion of important degrees of freedom, hyperons and kaons, in one single realistic prescription for dense matter in reasonable agreement with observations, there are several points for improvement. So far, we have not combined other baryons, such as $\Delta$ baryons, in our analysis of kaon condensation. These appear at similar densities as the hyperons and kaons and could modify our conclusions (see, e.g., the work in Ref.~\cite{Thapa:2021kfo,Ma:2022knr,Kaur:2024cfm}). We have also only studied s-wave condensation for kaons and have not allowed for pion condensation. Work on all these fronts is underway. 

\section{Acknowledgments}

We thank Khandker Quader for useful discussions and Ralf Rapp for guidance on the development of the mCMF model.
This material is based upon work supported by the National Science Foundation under grants No. PHY-2208724, PHY-2116686 and PHY-2514763, and within the framework of the MUSES collaboration, under Grant No. OAC-2103680. This material is also based upon work supported by the U.S. Department of Energy, Office of Science, Office of Nuclear Physics, under Award Numbers DE-SC0024700 and DE-SC0022023, as well as by the National Aeronautics and Space Agency (NASA) under Award Number 80NSSC24K0767.

\appendix
\section{Particle multiplets}

\begin{itemize}
    \item Baryon matrix
    \[
B=\begin{pmatrix}\frac{\Sigma^{0}}{\sqrt{2}}+\frac{\Lambda_{0}}{\sqrt{6}} & \Sigma^{+} & p\\
\Sigma^{-} & \frac{-\Sigma^{0}}{\sqrt{2}}+\frac{\Lambda^{0}}{\sqrt{6}} & n\\
\Xi^{-} & \Xi^{0} & -2\frac{\Lambda^{0}}{\sqrt{6}}
\end{pmatrix}\,.
\]

\item Scalar matrix 

\begin{align}
X=\begin{pmatrix}\frac{\delta^{0}+\sigma}{\sqrt{2}} & \delta^{+} & \kappa^{+}\\
\delta^{-} & \frac{-\delta^{0}+\sigma}{\sqrt{2}} & \kappa^{0}\\
\kappa^{-} & \bar{\kappa}^{0} & \zeta
\end{pmatrix}\,.
\end{align}

\item Scalar matrix after mean-field approximation (only temporal fields)

\begin{align}
X=\begin{pmatrix}\frac{\delta^{0}+\sigma}{\sqrt{2}} & 0 & 0\\
0 & \frac{-\delta^{0}+\sigma}{\sqrt{2}} & 0\\
0 & 0 & \zeta
\end{pmatrix}\,.
\end{align}

\item Vacuum expectation of scalar matrix after mean-field approximation

\begin{align}
\left\langle X \right\rangle=\begin{pmatrix}\frac{\sigma_0}{\sqrt{2}} & 0 & 0\\
0& \frac{\sigma_0}{\sqrt{2}} & 0\\
0& 0& \zeta_0
\end{pmatrix}\,,
\end{align}

where $\delta^0_0$=0.

\item Pseudoscalar octet matrix
 
\begin{align}
&P=\frac{\pi^{a}\left(x\right)\lambda_{a}}{\sqrt{2}}=\nonumber\\ &\begin{pmatrix}\frac{1}{\sqrt{2}}\left(\pi^{0}+\frac{\eta^{8}}{\sqrt{1+2w^{2}}}\right) & \pi^{+} & 2\frac{K^{+}}{w+1}\\
\pi^{-} & \frac{1}{\sqrt{2}}\left(-\pi^{0}+\frac{\eta^{8}}{\sqrt{1+2w^{2}}}\right) & 2\frac{K^{0}}{w+1}\\
2\frac{K^{-}}{w+1} & 2\frac{\bar{K^{0}}}{w+1} & -\sqrt{\frac{2}{1+2w^{2}}}\eta^{8}
\end{pmatrix},
\end{align}

where $w=\sqrt{2}\zeta_{0}/\sigma_{0}$. 

\item $A_p$ matrix
 
\begin{align}
A_p=\frac{1}{\sqrt{2}}\begin{pmatrix}m_\pi^2f_\pi & 0 & 0\\
0 & m_\pi^2f_\pi & 0\\
0 & 0 & 2m_K^2f_K-m_\pi^2f_\pi
\end{pmatrix}\,.
\end{align}

\end{itemize}

\clearpage

\bibliography{inspire}

@article{Kaplan:1986yq,
    author = "Kaplan, D. B. and Nelson, A. E.",
    title = "{Strange Goings on in Dense Nucleonic Matter}",
    doi = "10.1016/0370-2693(86)90331-X",
    journal = "Phys. Lett. B",
    volume = "175",
    pages = "57--63",
    year = "1986"
}

@article{Kaplan:1987sc,
    author = "Kaplan, D. B. and Nelson, A. E.",
    editor = "Speth, J.",
    title = "{Kaon Condensation in Dense Matter}",
    reportNumber = "ITP-834-STANFORD, HUTP-87-A057",
    doi = "10.1016/0375-9474(88)90442-3",
    journal = "Nucl. Phys. A",
    volume = "479",
    pages = "273c",
    year = "1988"
}

@article{Barnes:1992ca,
    author = "Barnes, Ted and Swanson, E. S.",
    title = "{Kaon - nucleon scattering amplitudes and Z* enhancements from quark Born diagrams}",
    eprint = "nucl-th/9212008",
    archivePrefix = "arXiv",
    reportNumber = "MIT-CTP-2169, ORNL-CCIP-92-15",
    doi = "10.1103/PhysRevC.49.1166",
    journal = "Phys. Rev. C",
    volume = "49",
    pages = "1166--1184",
    year = "1994"
}

@article{Mishra:2004te,
    author = "Mishra, A. and Bratkovskaya, E. L. and Schaffner-Bielich, J. and Schramm, S. and Stoecker, Horst",
    title = "{Kaons and antikaons in hot and dense hadronic matter}",
    eprint = "nucl-th/0402062",
    archivePrefix = "arXiv",
    doi = "10.1103/PhysRevC.70.044904",
    journal = "Phys. Rev. C",
    volume = "70",
    pages = "044904",
    year = "2004"
}

@article{Thorsson:1993bu,
    author = "Thorsson, Vesteinn and Prakash, Madappa and Lattimer, James M.",
    title = "{Composition, structure and evolution of neutron stars with kaon condensates}",
    eprint = "nucl-th/9305006",
    archivePrefix = "arXiv",
    reportNumber = "NORDITA-93-29-N, SUNY-NTG-92-33",
    doi = "10.1016/0375-9474(94)90407-3",
    journal = "Nucl. Phys. A",
    volume = "572",
    pages = "693--731",
    year = "1994",
    note = "[Erratum: Nucl.Phys.A 574, 851 (1994)]"
}

@article{Glendenning:1997ak,
    author = "Glendenning, Norman K. and Schaffner-Bielich, Jurgen",
    title = "{First order kaon condensate}",
    eprint = "astro-ph/9810290",
    archivePrefix = "arXiv",
    reportNumber = "LBNL-42330, LBL-42330",
    doi = "10.1103/PhysRevC.60.025803",
    journal = "Phys. Rev. C",
    volume = "60",
    pages = "025803",
    year = "1999"
}

@article{Kumar:2020vys,
    author = "Kumar, Rajesh and Kumar, Arvind",
    title = "{$\phi$ meson mass and decay width in strange hadronic matter}",
    eprint = "2005.05133",
    archivePrefix = "arXiv",
    primaryClass = "hep-ph",
    doi = "10.1103/PhysRevC.102.045206",
    journal = "Phys. Rev. C",
    volume = "102",
    number = "4",
    pages = "045206",
    year = "2020"
}

@article{Kumari:2022jvq,
    author = "Kumari, Manisha and Kumar, Arvind",
    title = "{Antikaon condensation in magnetized neutron stars}",
    doi = "10.1142/S0218301322500501",
    journal = "Int. J. Mod. Phys. E",
    volume = "31",
    number = "05",
    pages = "2250050",
    year = "2022"
}

@article{Antoniadis:2013pzd,
    author = "Antoniadis, John and others",
    title = "{A Massive Pulsar in a Compact Relativistic Binary}",
    eprint = "1304.6875",
    archivePrefix = "arXiv",
    primaryClass = "astro-ph.HE",
    doi = "10.1126/science.1233232",
    journal = "Science",
    volume = "340",
    pages = "6131",
    year = "2013"
}

@article{Fonseca:2021wxt,
    author = "Fonseca, E. and others",
    title = "{Refined Mass and Geometric Measurements of the High-mass PSR J0740+6620}",
    eprint = "2104.00880",
    archivePrefix = "arXiv",
    primaryClass = "astro-ph.HE",
    doi = "10.3847/2041-8213/ac03b8",
    journal = "Astrophys. J. Lett.",
    volume = "915",
    number = "1",
    pages = "L12",
    year = "2021"
}

@article{Chamel:2013efa,
    author = "Chamel, N. and Haensel, P. and Zdunik, J. L. and Fantina, A. F.",
    title = "{On the Maximum Mass of Neutron Stars}",
    eprint = "1307.3995",
    archivePrefix = "arXiv",
    primaryClass = "astro-ph.HE",
    doi = "10.1142/S021830131330018X",
    journal = "Int. J. Mod. Phys. E",
    volume = "22",
    pages = "1330018",
    year = "2013"
}

@article{Kumar:2025rxj,
    author = "Kumar, Rajesh and Grefa, Joaquin and Maslov, Konstantin and Wang, Yuhan and Kumar, Arvind and Rapp, Ralf and Ratti, Claudia and Dexheimer, Veronica",
    title = "{Interacting mesons as degrees of freedom in a chiral model}",
    eprint = "2503.03057",
    archivePrefix = "arXiv",
    primaryClass = "nucl-th",
    doi = "10.1103/PhysRevD.111.074029",
    journal = "Phys. Rev. D",
    volume = "111",
    number = "7",
    pages = "074029",
    year = "2025"
}

@article{LIGOScientific:2018hze,
    author = "Abbott, B. P. and others",
    collaboration = "LIGO Scientific, Virgo",
    title = "{Properties of the binary neutron star merger GW170817}",
    eprint = "1805.11579",
    archivePrefix = "arXiv",
    primaryClass = "gr-qc",
    doi = "10.1103/PhysRevX.9.011001",
    journal = "Phys. Rev. X",
    volume = "9",
    number = "1",
    pages = "011001",
    year = "2019"
}

@article{Dittmann:2024mbo,
    author = "Dittmann, Alexander J. and others",
    title = "{A More Precise Measurement of the Radius of PSR J0740+6620 Using Updated NICER Data}",
    eprint = "2406.14467",
    archivePrefix = "arXiv",
    primaryClass = "astro-ph.HE",
    doi = "10.3847/1538-4357/ad5f1e",
    journal = "Astrophys. J.",
    volume = "974",
    number = "2",
    pages = "295",
    year = "2024"
}

@article{Choudhury:2024xbk,
    author = "Choudhury, Devarshi and others",
    title = "{A NICER View of the Nearest and Brightest Millisecond Pulsar: PSR J0437{\textendash}4715}",
    eprint = "2407.06789",
    archivePrefix = "arXiv",
    primaryClass = "astro-ph.HE",
    doi = "10.3847/2041-8213/ad5a6f",
    journal = "Astrophys. J. Lett.",
    volume = "971",
    number = "1",
    pages = "L20",
    year = "2024"
}

@article{Lee:1994my,
    author = "Lee, Chang-Hwan and Jung, Hong and Min, Dong-Pil and Rho, Mannque",
    title = "{Kaon - nucleon scattering from chiral Lagrangians}",
    eprint = "hep-ph/9401245",
    archivePrefix = "arXiv",
    reportNumber = "SNUTP-93-81",
    doi = "10.1016/0370-2693(94)91185-1",
    journal = "Phys. Lett. B",
    volume = "326",
    pages = "14--20",
    year = "1994"
}

@article{Lattimer:1991ib,
    author = "Lattimer, J. M. and Prakash, M. and Pethick, C. J. and Haensel, P.",
    title = "{Direct URCA process in neutron stars}",
    doi = "10.1103/PhysRevLett.66.2701",
    journal = "Phys. Rev. Lett.",
    volume = "66",
    pages = "2701--2704",
    year = "1991"
}

@article{Page:2005fq,
    author = "Page, Dany and Geppert, Ulrich and Weber, Fridolin",
    title = "{The Cooling of compact stars}",
    eprint = "astro-ph/0508056",
    archivePrefix = "arXiv",
    doi = "10.1016/j.nuclphysa.2005.09.019",
    journal = "Nucl. Phys. A",
    volume = "777",
    pages = "497--530",
    year = "2006"
}

@article{Gudmundsson1983,
author = {Gudmundsson, E. H. and Pethick, C. J. and Epstein, R. I.},
doi = {10.1086/161292},
issn = {0004-637X},
journal = {The Astrophysical Journal},
month = {sep},
pages = {286},
title = {{Structure of neutron star envelopes}},
url = {http://adsabs.harvard.edu/doi/10.1086/161292},
volume = {272},
year = {1983}
}

@article{Malik:2021nas,
    author = "Malik, Tuhin and Banik, Sarmistha and Bandyopadhyay, Debades",
    title = "{Equation-of-state Table with Hyperon and Antikaon for Supernova and Neutron Star Merger}",
    eprint = "2104.00775",
    archivePrefix = "arXiv",
    primaryClass = "astro-ph.HE",
    doi = "10.3847/1538-4357/abe860",
    journal = "Astrophys. J.",
    volume = "910",
    number = "2",
    pages = "96",
    year = "2021"
}

@article{Thapa:2021kfo,
    author = "Thapa, Vivek Baruah and Sinha, Monika and Li, Jia Jie and Sedrakian, Armen",
    title = "{Massive $\Delta$-resonance admixed hypernuclear stars with antikaon condensations}",
    eprint = "2102.08787",
    archivePrefix = "arXiv",
    primaryClass = "astro-ph.HE",
    doi = "10.1103/PhysRevD.103.063004",
    journal = "Phys. Rev. D",
    volume = "103",
    number = "6",
    pages = "063004",
    year = "2021"
}

@article{Ma:2022fmu,
    author = "Ma, Fu and Guo, Wenjun and Wu, Chen",
    title = "{Kaon meson condensate in neutron star matter including hyperons}",
    eprint = "2202.03001",
    archivePrefix = "arXiv",
    primaryClass = "nucl-th",
    doi = "10.1103/PhysRevC.105.015807",
    journal = "Phys. Rev. C",
    volume = "105",
    number = "1",
    pages = "015807",
    year = "2022"
}

@article{Sharifi:2021gvi,
    author = "Sharifi, Z. and Bigdeli, M. and Alvarez-Castillo, D. and Nasiri, E.",
    title = "{Binary neutron star mergers within kaon condensation: GW170817}",
    eprint = "2112.09730",
    archivePrefix = "arXiv",
    primaryClass = "astro-ph.HE",
    doi = "10.1088/1402-4896/ac30a5",
    journal = "Phys. Scripta",
    volume = "96",
    number = "12",
    pages = "125311",
    year = "2021"
}

@article{Panda:2013oza,
    author = "Panda, Prafulla K. and Menezes, D{\'e}bora P. and Provid{\^e}ncia, Constan{\c{c}}a",
    title = "{Effects of the symmetry energy on the kaon condensates in the quark-meson coupling model}",
    eprint = "1311.2739",
    archivePrefix = "arXiv",
    primaryClass = "nucl-th",
    doi = "10.1103/PhysRevC.89.045803",
    journal = "Phys. Rev. C",
    volume = "89",
    number = "4",
    pages = "045803",
    year = "2014"
}

@ARTICLE{1982ApJ...259L..19G,
       author = {{Gudmundsson}, E.~H. and {Pethick}, C.~J. and {Epstein}, R.~I.},
        title = "{Neutron star envelopes}",
      journal = {Astrophys. J. Lett.},
         year = 1982,
        month = aug,
       volume = {259},
        pages = {L19-L23},
          doi = {10.1086/183840},
       adsurl = {https://ui.adsabs.harvard.edu/abs/1982ApJ...259L..19G}
}

@article{Yakovlev:2004iq,
    author = "Yakovlev, Dima G. and Pethick, C. J.",
    title = "{Neutron star cooling}",
    eprint = "astro-ph/0402143",
    archivePrefix = "arXiv",
    doi = "10.1146/annurev.astro.42.053102.134013",
    journal = "Ann. Rev. Astron. Astrophys.",
    volume = "42",
    pages = "169--210",
    year = "2004"
}

@article{Yakovlev:2000jp,
    author = "Yakovlev, D. G. and Kaminker, A. D. and Gnedin, Oleg Y. and Haensel, P.",
    title = "{Neutrino emission from neutron stars}",
    eprint = "astro-ph/0012122",
    archivePrefix = "arXiv",
    doi = "10.1016/S0370-1573(00)00131-9",
    journal = "Phys. Rept.",
    volume = "354",
    pages = "1",
    year = "2001"
}

@article{Schaab:1996jm,
    author = "Schaab, Christoph and Weber, Fridolin and Weigel, Manfred K. and Glendenning, Norman K.",
    title = "{Thermal evolution of compact stars}",
    eprint = "astro-ph/9603142",
    archivePrefix = "arXiv",
    doi = "10.1016/0375-9474(96)00164-9",
    journal = "Nucl. Phys. A",
    volume = "605",
    pages = "531",
    year = "1996"
}

@book{Weber:1999qn,
    author = "Weber, F.",
    title = "{Pulsars as astrophysical laboratories for nuclear and particle physics}",
    doi = "10.1201/9780203741719",
    isbn = "978-0-203-74171-9",
    year = "1999"
}

@ARTICLE{1990ApJ...354L..17P,
       author = {{Page}, Dany and {Baron}, E.},
        title = "{Strangeness, Condensation, Nucleon Superfluidity, and Cooling of Neutron Stars}",
      journal = {"Astrophys. J. Let."},
         year = 1990,
        month = may,
       volume = {354},
        pages = {L17},
          doi = {10.1086/185712},
       adsurl = {https://ui.adsabs.harvard.edu/abs/1990ApJ...354L..17P}
}

@article{Friedman:1998xa,
    author = "Friedman, E. and Gal, A. and Mares, J.",
    title = "{K- Nucleus relativistic mean field potential consistent with kaonic atoms}",
    eprint = "nucl-th/9804072",
    archivePrefix = "arXiv",
    doi = "10.1103/PhysRevC.60.024314",
    journal = "Phys. Rev. C",
    volume = "60",
    pages = "024314",
    year = "1999"
}

@article{Friedman:1994hx,
    author = "Friedman, E. and Gal, A. and Batty, C. J.",
    title = "{Density dependent K- nuclear optical potentials from kaonic atoms}",
    doi = "10.1016/0375-9474(94)90921-0",
    journal = "Nucl. Phys. A",
    volume = "579",
    pages = "518--538",
    year = "1994"
}

@article{Ramos:1999ku,
    author = "Ramos, A. and Oset, E.",
    title = "{The Properties of anti-K in the nuclear medium}",
    eprint = "nucl-th/9906016",
    archivePrefix = "arXiv",
    doi = "10.1016/S0375-9474(99)00846-5",
    journal = "Nucl. Phys. A",
    volume = "671",
    pages = "481--502",
    year = "2000"
}

@article{Mai:2012dt,
    author = "Mai, Maxim and Meissner, Ulf-G.",
    title = "{New insights into antikaon-nucleon scattering and the structure of the Lambda(1405)}",
    eprint = "1202.2030",
    archivePrefix = "arXiv",
    primaryClass = "nucl-th",
    doi = "10.1016/j.nuclphysa.2013.01.032",
    journal = "Nucl. Phys. A",
    volume = "900",
    pages = "51 -- 64",
    year = "2013"
}

@article{Ikeda:2012au,
    author = "Ikeda, Yoichi and Hyodo, Tetsuo and Weise, Wolfram",
    title = "{Chiral SU(3) theory of antikaon-nucleon interactions with improved threshold constraints}",
    eprint = "1201.6549",
    archivePrefix = "arXiv",
    primaryClass = "nucl-th",
    doi = "10.1016/j.nuclphysa.2012.01.029",
    journal = "Nucl. Phys. A",
    volume = "881",
    pages = "98--114",
    year = "2012"
}

@article{Meissner:2004jr,
    author = "Meissner, Ulf G. and Raha, Udit and Rusetsky, Akaki",
    title = "{Spectrum and decays of kaonic hydrogen}",
    eprint = "hep-ph/0402261",
    archivePrefix = "arXiv",
    reportNumber = "HISKP-TH-04-01",
    doi = "10.1140/epjc/s2004-01859-4",
    journal = "Eur. Phys. J. C",
    volume = "35",
    pages = "349--357",
    year = "2004"
}

@article{Meissner:2006gx,
    author = "Meissner, Ulf-G. and Raha, Udit and Rusetsky, Akaki",
    title = "{Kaon-nucleon scattering lengths from kaonic deuterium experiments}",
    eprint = "nucl-th/0603029",
    archivePrefix = "arXiv",
    reportNumber = "HISKP-TH-06-01, FZJ-IKP-TH-2006-05",
    doi = "10.1140/epjc/s2006-02578-6",
    journal = "Eur. Phys. J. C",
    volume = "47",
    pages = "473--480",
    year = "2006"
}

@article{Doring:2011xc,
    author = {D{\"o}ring, M. and Mei{\ss}ner, U. -G.},
    title = "{Kaon{\textendash}nucleon scattering lengths from kaonic deuterium experiments revisited}",
    eprint = "1108.5912",
    archivePrefix = "arXiv",
    primaryClass = "nucl-th",
    doi = "10.1016/j.physletb.2011.09.099",
    journal = "Phys. Lett. B",
    volume = "704",
    pages = "663--666",
    year = "2011"
}

@article{Borasoy:2006sr,
    author = "Borasoy, B. and Meissner, U. -G. and Nissler, R.",
    title = "{K- p scattering length from scattering experiments}",
    eprint = "hep-ph/0606108",
    archivePrefix = "arXiv",
    reportNumber = "HISKP-TH-06-15, FZJ-IKP-TH-2006-16",
    doi = "10.1103/PhysRevC.74.055201",
    journal = "Phys. Rev. C",
    volume = "74",
    pages = "055201",
    year = "2006"
}

@article{ALICE:2021szj,
    author = "Acharya, Shreyasi and others",
    collaboration = "ALICE",
    title = "{Kaon{\textendash}proton strong interaction at low relative momentum via femtoscopy in Pb{\textendash}Pb collisions at the LHC}",
    eprint = "2105.05683",
    archivePrefix = "arXiv",
    primaryClass = "nucl-ex",
    reportNumber = "CERN-EP-2021-080",
    doi = "10.1016/j.physletb.2021.136708",
    journal = "Phys. Lett. B",
    volume = "822",
    pages = "136708",
    year = "2021"
}

@ARTICLE{1988PhRvC..38.1010W,
       author = {{Wiringa}, R.~B. and {Fiks}, V. and {Fabrocini}, A.},
        title = "{Equation of state for dense nucleon matter}",
      journal = {\prc},
         year = 1988,
        month = aug,
       volume = {38},
       number = {2},
        pages = {1010-1037},
          doi = {10.1103/PhysRevC.38.1010},
       adsurl = {https://ui.adsabs.harvard.edu/abs/1988PhRvC..38.1010W}
}

@article{Zhang:2020wov,
    author = "Zhang, Nai-Bo and Li, Bao-An",
    title = "{Constraints on the muon fraction and density profile in neutron stars}",
    eprint = "2002.06446",
    archivePrefix = "arXiv",
    primaryClass = "astro-ph.HE",
    doi = "10.3847/1538-4357/ab7dbc",
    journal = "Astrophys. J.",
    volume = "893",
    pages = "61",
    year = "2020"
}

@article{Mishra:2009bp,
    author = "Mishra, Amruta and Kumar, Arvind and Sanyal, Sambuddha and Dexheimer, V. and Schramm, Stefan",
    title = "{Kaon properties in (proto)neutron stars}",
    eprint = "0905.3518",
    archivePrefix = "arXiv",
    primaryClass = "nucl-th",
    doi = "10.1140/epja/i2010-10986-x",
    journal = "Eur. Phys. J. A",
    volume = "45",
    pages = "169--177",
    year = "2010"
}

@article{osti_4718088,
  author       = {Baym, G and Pethick, C and Sutherland, P},
  title        = {The Ground State of Matter at High Densities: Equation of State and Stellar Models.},
  doi          = {10.1086/151216},
  url          = {https://www.osti.gov/biblio/4718088},
  journal      = {Astrophys. J. 170:  No. 2, 299-317(1 Dec 1971).},
  place        = {Country unknown/Code not available},
  year         = {1970},
  month        = {12}
}

@article{Hyodo:2007jq,
    author = "Hyodo, Tetsuo and Weise, Wolfram",
    title = "{Effective anti-K N interaction based on chiral SU(3) dynamics}",
    eprint = "0712.1613",
    archivePrefix = "arXiv",
    primaryClass = "nucl-th",
    doi = "10.1103/PhysRevC.77.035204",
    journal = "Phys. Rev. C",
    volume = "77",
    pages = "035204",
    year = "2008"
}

@article{Horowitz:2002mb,
    author = "Horowitz, C. J. and Piekarewicz, J.",
    title = "{Constraining URCA cooling of neutron stars from the neutron radius of Pb-208}",
    eprint = "nucl-th/0207067",
    archivePrefix = "arXiv",
    doi = "10.1103/PhysRevC.66.055803",
    journal = "Phys. Rev. C",
    volume = "66",
    pages = "055803",
    year = "2002"
}

@article{Dexheimer:2018dhb,
    author = "Dexheimer, Veronica and de Oliveira Gomes, Rosana and Schramm, Stefan and Pais, Helena",
    title = "{What do we learn about vector interactions from GW170817?}",
    eprint = "1810.06109",
    archivePrefix = "arXiv",
    primaryClass = "nucl-th",
    doi = "10.1088/1361-6471/ab01f0",
    journal = "J. Phys. G",
    volume = "46",
    number = "3",
    pages = "034002",
    year = "2019"
}

@article{Kumar:2024owe,
    author = "Kumar, Rajesh and Wang, Yuhan and Camacho, Nikolas Cruz and Kumar, Arvind and Noronha-Hostler, Jacquelyn and Dexheimer, Veronica",
    title = "{Modern nuclear and astrophysical constraints of dense matter in a redefined chiral approach}",
    eprint = "2401.12944",
    archivePrefix = "arXiv",
    primaryClass = "nucl-th",
    doi = "10.1103/PhysRevD.109.074008",
    journal = "Phys. Rev. D",
    volume = "109",
    number = "7",
    pages = "074008",
    year = "2024"
}

@article{Cruz-Camacho:2024odu,
    author = "Cruz-Camacho, Nikolas and Kumar, Rajesh and Reinke Pelicer, Mateus and Peterson, Jeff and Manning, T. Andrew and Haas, Roland and Dexheimer, Veronica and Noronha-Hostler, Jaquelyn",
    collaboration = "MUSES",
    title = "{Phase stability in the three-dimensional open-source code for the chiral mean-field model}",
    eprint = "2409.06837",
    archivePrefix = "arXiv",
    primaryClass = "nucl-th",
    doi = "10.1103/PhysRevD.111.094030",
    journal = "Phys. Rev. D",
    volume = "111",
    number = "9",
    pages = "094030",
    year = "2025"
}

@misc{S:2025xfm,
    author = "S., Athira and Sinha, Monika and Bandyopadhyay, Debades and Thapa, Vivek Baruah and Parmar, Vishal",
    title = "{Antikaon condensed dense matter in neutron star with SU(3) flavour symmetry}",
    eprint = "2504.06859",
    archivePrefix = "arXiv",
    primaryClass = "nucl-th",
    month = "4",
    year = "2025"
}

@article{Kundu:2022nva,
    author = "Kundu, Debraj and Thapa, Vivek Baruah and Sinha, Monika",
    title = "{(Anti)kaon condensation in strongly magnetized dense matter}",
    eprint = "2210.14565",
    archivePrefix = "arXiv",
    primaryClass = "astro-ph.HE",
    doi = "10.1103/PhysRevC.107.035807",
    journal = "Phys. Rev. C",
    volume = "107",
    number = "3",
    pages = "035807",
    year = "2023"
}

@article{Ma:2022knr,
    author = "Ma, Fu and Wu, Chen and Guo, Wenjun",
    title = "{Kaon-meson condensation and {\ensuremath{\Delta}} resonance in hyperonic stellar matter within a relativistic mean-field model}",
    eprint = "2211.11498",
    archivePrefix = "arXiv",
    primaryClass = "nucl-th",
    doi = "10.1103/PhysRevC.107.045804",
    journal = "Phys. Rev. C",
    volume = "107",
    number = "4",
    pages = "045804",
    year = "2023"
}

@article{Kaur:2024cfm,
    author = "Kaur, Manpreet and Kumar, Arvind",
    title = "{Kaons and antikaons in isospin asymmetric dense resonance matter at finite temperature}",
    eprint = "2410.15685",
    archivePrefix = "arXiv",
    primaryClass = "hep-ph",
    doi = "10.1103/PhysRevD.110.114054",
    journal = "Phys. Rev. D",
    volume = "110",
    number = "11",
    pages = "114054",
    year = "2024"
}

@article{Parmar:2024qff,
    author = "Parmar, Vishal and Thapa, Vivek Baruah and Kumar, Anil and Bandyopadhyay, Debades and Sinha, Monika",
    title = "{Bayesian inference of the dense-matter equation of state of neutron stars with antikaon condensation}",
    eprint = "2409.19451",
    archivePrefix = "arXiv",
    primaryClass = "astro-ph.HE",
    doi = "10.1103/PhysRevC.110.045804",
    journal = "Phys. Rev. C",
    volume = "110",
    number = "4",
    pages = "045804",
    year = "2024"
}

@article{GuhaRoy:2025kht,
    author = "Guha Roy, Debanjan and Banik, Sarmistha",
    title = "{Signatures of K{\ensuremath{-}} condensation on neutron star structure and f{\ensuremath{-}}mode frequencies}",
    eprint = "2509.15376",
    archivePrefix = "arXiv",
    primaryClass = "nucl-th",
    doi = "10.1016/j.physletb.2025.139990",
    journal = "Phys. Lett. B",
    volume = "871",
    pages = "139990",
    year = "2025"
}

@article{Thakur:2025axg,
    author = "Thakur, Prashant and Kumaran, Yashmitha and Sudarsan, Lakshana and Kunnampully, Krishna and Sharma, B. K. and Jha, T. K.",
    title = "{Implications of the {\ensuremath{\sigma}}-cut potential on antikaon condensates in neutron stars}",
    eprint = "2502.18882",
    archivePrefix = "arXiv",
    primaryClass = "astro-ph.HE",
    doi = "10.1103/PhysRevC.111.035801",
    journal = "Phys. Rev. C",
    volume = "111",
    number = "3",
    pages = "035801",
    year = "2025"
}

@article{Muto:2025jaq,
    author = "Muto, Takumi and Maruyama, Toshiki and Tatsumi, Toshitaka",
    title = "{Chiral Symmetry in Dense Matter with Meson Condensation}",
    eprint = "2502.12192",
    archivePrefix = "arXiv",
    primaryClass = "nucl-th",
    doi = "10.3390/sym17020270",
    journal = "Symmetry",
    volume = "17",
    number = "2",
    pages = "270",
    year = "2025"
}

@article{Muto:2024upf,
    author = "Muto, Takumi",
    title = "{Properties of a kaon-condensed phase in hyperon-mixed matter with three-baryon forces}",
    eprint = "2411.09967",
    archivePrefix = "arXiv",
    primaryClass = "nucl-th",
    doi = "10.1103/PhysRevC.111.045802",
    journal = "Phys. Rev. C",
    volume = "111",
    number = "4",
    pages = "045802",
    year = "2025"
}

@article{Muto:2021jms,
    author = "Muto, Takumi and Maruyama, Toshiki and Tatsumi, Toshitaka",
    title = "{Effects of three-baryon forces on kaon condensation in hyperon-mixed matter}",
    eprint = "2106.03449",
    archivePrefix = "arXiv",
    primaryClass = "nucl-th",
    doi = "10.1016/j.physletb.2021.136587",
    journal = "Phys. Lett. B",
    volume = "820",
    pages = "136587",
    year = "2021"
}

@article{Muto:2022ces,
    author = "Muto, Takumi and Maruyama, Toshiki and Tatsumi, Toshitaka",
    title = "{Kaon{\textendash}baryon coupling schemes and kaon condensation in hyperon-mixed matter}",
    eprint = "2207.00242",
    archivePrefix = "arXiv",
    primaryClass = "nucl-th",
    doi = "10.1093/ptep/ptac115",
    journal = "PTEP",
    volume = "2022",
    number = "9",
    pages = "093D03",
    year = "2022"
}

@article{Demorest:2010bx,
    author = "Demorest, Paul and Pennucci, Tim and Ransom, Scott and Roberts, Mallory and Hessels, Jason",
    title = "{Shapiro Delay Measurement of A Two Solar Mass Neutron Star}",
    eprint = "1010.5788",
    archivePrefix = "arXiv",
    primaryClass = "astro-ph.HE",
    doi = "10.1038/nature09466",
    journal = "Nature",
    volume = "467",
    pages = "1081--1083",
    year = "2010"
}

@article{Lee:1994jj,
    author = "Lee, Chang-Hwan and Brown, G. E. and Min, Dong-Pil and Rho, Mannque",
    title = "{An Effective chiral Lagrangian approach to kaon - nuclear interactions: Kaonic atom and kaon condensation}",
    eprint = "hep-ph/9406311",
    archivePrefix = "arXiv",
    reportNumber = "SNUTP-94-50",
    doi = "10.1016/0375-9474(94)00623-U",
    journal = "Nucl. Phys. A",
    volume = "585",
    pages = "401--449",
    year = "1995"
}

@article{Banik:2001yw,
    author = "Banik, Sarmistha and Bandyopadhyay, Debades",
    title = "{A Third family of superdense stars in the presence of anti-kaon condensates}",
    eprint = "astro-ph/0106406",
    archivePrefix = "arXiv",
    doi = "10.1103/PhysRevC.64.055805",
    journal = "Phys. Rev. C",
    volume = "64",
    pages = "055805",
    year = "2001"
}

@article{Brown:1993yv,
    author = "Brown, G. E. and Lee, Chang-Hwan and Rho, Mannque and Thorsson, Vesteinn",
    title = "{From kaon - nuclear interactions to kaon condensation}",
    eprint = "hep-ph/9304204",
    archivePrefix = "arXiv",
    reportNumber = "NORDITA-93-30-N, SUNY-NTG-93-7",
    doi = "10.1016/0375-9474(94)90335-2",
    journal = "Nucl. Phys. A",
    volume = "567",
    pages = "937--956",
    year = "1994"
}

@article{Schaffner:1995th,
    author = "Schaffner, Jurgen and Mishustin, Igor N.",
    title = "{Hyperon rich matter in neutron stars}",
    eprint = "nucl-th/9506011",
    archivePrefix = "arXiv",
    doi = "10.1103/PhysRevC.53.1416",
    journal = "Phys. Rev. C",
    volume = "53",
    pages = "1416--1429",
    year = "1996"
}

@article{Mishra:2006wy,
    author = "Mishra, Amruta and Schramm, Stefan",
    title = "{Isospin dependent kaon and antikaon optical potentials in dense hadronic matter}",
    eprint = "nucl-th/0607050",
    archivePrefix = "arXiv",
    doi = "10.1103/PhysRevC.74.064904",
    journal = "Phys. Rev. C",
    volume = "74",
    pages = "064904",
    year = "2006"
}

\end{document}